\documentclass[a4paper,fleqn,usenatbib]{mnras}

\usepackage{newtxtext,newtxmath}

\usepackage[T1]{fontenc}
\usepackage{ae,aecompl}

\usepackage{graphicx}	
\usepackage{amsmath}	

\usepackage{longtable}

\usepackage{pdflscape}
\usepackage{multicol}

\title[Exploring AQ Men tilted disc]{Exploring the tilted accretion disc of AQ Men with TESS}

\author[I{\l}kiewicz et al.]{Krystian I{\l}kiewicz,$^{1,2}$\thanks{E-mail: krystian.a.ilkiewicz@durham.ac.uk } Simone Scaringi,$^{1,2}$ James M.C. Court,$^{2}$ Thomas J. Maccarone,$^{2}$
\newauthor Diego Altamirano,$^{3}$ Corey W. Bradshaw,$^{2}$ Nathalie Degenaar,$^{4}$ Matteo Fratta,$^{1,2}$ 
 \newauthor Colin Littlefield,$^{5}$ Tariq Shahbaz,$^{6,7}$ Rudy Wijnands$^{4}$
\\
$^{1}$Centre for Extragalactic Astronomy, Department of Physics, University of Durham, South Road, Durham, DH1 3LE, UK\\
$^{2}$Department of Physics and Astronomy, Texas Tech University, PO Box 41051, Lubbock, TX 79409, USA\\
$^{3}$School of Physics and Astronomy, University of Southampton, Southampton SO17 1BJ, UK\\
$^{4}$Anton Pannekoek Institute for Astronomy, University of Amsterdam, Postbus 94249, 1090 GE Amsterdam, The Netherlands\\
$^{5}$Department of Physics, University of Notre Dame, Notre Dame, IN 46556, USA\\
$^{6}$Instituto de Astrof\'\i{}sica de Canarias (IAC), E-38205 La Laguna, Tenerife, Spain\\
$^{7}$Departamento de  Astrof\'\i{}sica, Universidad de La Laguna (ULL), E-38206 La Laguna, Tenerife, Spain \\
}

\date{Accepted XXX. Received YYY; in original form ZZZ}

\pubyear{2020}

\begin{document}
\label{firstpage}
\pagerange{\pageref{firstpage}--\pageref{lastpage}}
\maketitle

\begin{abstract}

AQ Men is a nova-like variable which is presumed to have a tilted, precessing accretion disc. Grazing eclipses in this system have been speculated to be useful in exploring the geometry of its accretion disc. In this work we analysed \textit{TESS} observations of AQ Men, which provide the best light curve of this object thus far. We show that the depths of the eclipses are changing with the orientation of the accretion disc, which means that they can serve as a direct test of the tilted accretion disc models. The precession period of the accretion disc is increasing during the \textit{TESS} observations. However, it is still shorter than the period determined in the previous studies. The amplitude of the variability related to the precession of the accretion disc varies, and so does the shape of this variability. Moreover, we have detected a positive superhump that was previously unseen in AQ Men. Interestingly, the positive superhump has a strongly non-sinusoidal shape, which is not expected for a nova-like variable.

\end{abstract}




\section{Introduction}

Cataclysmic variables are binary systems in which a late-type star is transferring mass to a white dwarf via Roche lobe overflow. In these systems, the mass is usually accreted through an accretion disc, unless the white dwarf has a high magnetic field, in which case matter is accreted directly along field lines onto the magnetic poles. In dwarf novae, the mass accretion rate is relatively low and the accretion disc has a temperature suitable to trigger thermal instability, causing repeated dwarf nova outbursts. Nova-like variables, on the other hand, are systems in which no dwarf nova or classical nova outbursts have been recorded. Nova-like variables often have a high mass-transfer rate. Reviews of cataclysmic variables are presented in \citet{2003cvs..book.....W} and \citet{2011ApJS..194...28K}.

Due to their relative simplicity, cataclysmic variables with low magnetic fields are excellent laboratories to study accretion discs. One of the more puzzling phenomena related to accretion discs in cataclysmic variables are superhumps.  Positive and negative superhumps are quasiperiodic signals that have periods a few percent longer or shorter compared to the orbital period, respectively. Both positive and negative superhumps can occur individually or simultaneously  \citep[e.g.][]{1999dicb.conf...61P}.  Positive superhumps are produced when a tidal instability causes elongation of the accretion disc \citep{1988MNRAS.232...35W,1988MNRAS.233..529W}. It has been shown that a 3:1 tidal resonance in the accretion disc is the reason behind this instability \citep{1990PASJ...42..135H}. A positive superhump is understood as being a beat frequency of a prograde apsida precession of an elliptical disc, and the orbital frequency of the binary system \citep{1998ApJ...506..360S,2001MNRAS.324..529R,2011ApJ...741..105W}. While the origin of negative superhumps is not well understood, they are thought to be a beat frequency of a shifting hotspot around the face of a tilted, retrogradely precessing disc and the orbital frequency  \citep{2000ApJ...535L..39W,2009MNRAS.398.2110W}. In addition, magnetic fields have been suggested as a significant factor leading to negative superhumps by lifting the accretion disc out of the orbital plane \citep{2015ApJ...803...55T}. In some cases, a superorbital signal is detected together with a negative superhump. The frequency of the superorbital signal is always equal to the difference between the orbital and negative superhump frequencies. This implies that the superorbital period is the period of the accretion disc precession.

\begin{figure}
\resizebox{\hsize}{!}{   \includegraphics{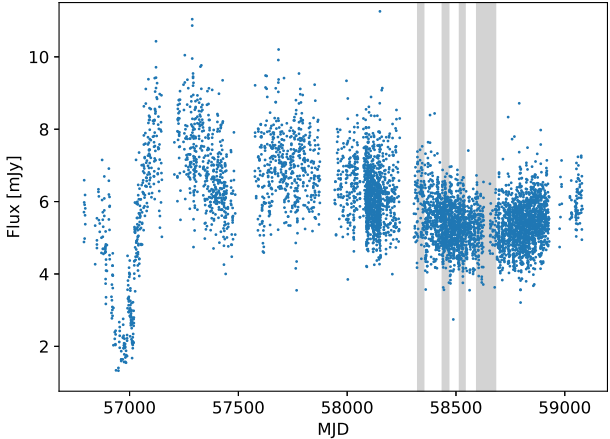}}
\caption{ASAS-SN light-curve of AQ~Men. The times of the \textit{TESS} sectors in which AQ~Men was observed are marked with grey. }
\label{fig:ASASSN}
\end{figure}

AQ~Men (= EC 0511-7955) was classified as a blue object in the Edinburgh-Cape Blue Object Survey \citep{1995Ap&SS.230..101S}. During a spectroscopic  follow-up,  \citet{2001MNRAS.325...89C}  identified AQ~Men as a cataclysmic variable based on the presence of broad, double-peaked Balmer emission lines. While a low excitation spectrum and high amplitude of flickering in AQ~Men led \citet{2001MNRAS.325...89C} to suggest that this object could be a dwarf nova, the authors also noted that the presence of a \mbox{C\,{\sc iii}}/\mbox{N\,{\sc iii}\,4650} blend in emission is more consistent with a nova-like variable. Similarly, \citet{2009ApJ...701.1091G} showed that the UV spectrum is most consistent with a dwarf nova in quiescence, but nova-like classification cannot be ruled out. However,  based on spectral features of AQ~Men, \citet{2012MmSAI..83..610S} classified this object as a SW~Sex star, a subclass of nova-like variables. Moreover, \citet{2020NewA...7801369B} noted that the absolute brightness of AQ~Men is too high for a dwarf nova, while it is consistent with a faint nova-like variable. Since no dwarf nova outburst has been observed in AQ~Men thus far, the nova-like classification is more likely \citep{2013MNRAS.435..707A}. In order to determine the nature of AQ~Men, we extracted a ASAS-SN light-curve of this object (\citealt{2014ApJ...788...48S,2017PASP..129j4502K}). In the light-curve a low state is visible at MJD$\simeq$56970 that lasted $\sim$100~days (Fig.~\ref{fig:ASASSN}). Using the presence of a low state we unambiguously classify AQ~Men as a nova-like variable.

While the photometric observations of AQ~Men \citet{2001MNRAS.325...89C} did not detect periodic signals, the authors suggested an orbital period of 0.130$\pm$0.014\,d based on radial velocity measurements.  The orbital period was later refined to 0.141\,d (or 7.0686\,c/d) by \citet{2013MNRAS.435..707A}, where it was shown that the system is eclipsing. In addition to the orbital period, \citet{2013MNRAS.435..707A} detected a signal at 0.263\,c/d, as well as a weaker signal at  7.332\,c/d.  These signals were interpreted by authors as superorbital and negative superhump periods, respectively. However, neither superorbital nor negative superhump periods were detected by \citet{2020NewA...7801369B};  the only signals detected by that author were a relatively strong signal at 8.25\,d (or 0.12\,c/d) and a weak signal at the orbital period proposed by \citet{2013MNRAS.435..707A}.

Both the data used by \citet{2013MNRAS.435..707A} and \citet{2020NewA...7801369B} were obtained by ground-based telescopes. While \citet{2013MNRAS.435..707A} observed AQ~Men for 33 nights and \citet{2020NewA...7801369B} only for 8 nights, the data collected by \citet{2020NewA...7801369B} should be enough to yield consistent results to the \citet{2013MNRAS.435..707A} study. In order to resolve the issue of variable periods present in AQ~Men we employed \textit{TESS} observations, which result in much more sensitive and longer monitoring of the object compared to previous studies. We present these observations in Section~\ref{sec:obs}.  In Section~\ref{sec:res}, we present the detection of negative and positive superhumps.  In Section~\ref{sec:evol}, we discuss the evolution of superhumps, as well as the variability of the superorbital period of this system. In Section~\ref{sec:supvar} we present changes in positive and negative superhumps, as well as in the depth of eclipses, that are related to the precession of the accretion disc.  Finally, we give a summary of our results in Section~\ref{sec:conclusions}.

\section{Observations}\label{sec:obs}

\begin{figure*}
\resizebox{\hsize}{!}{   \includegraphics{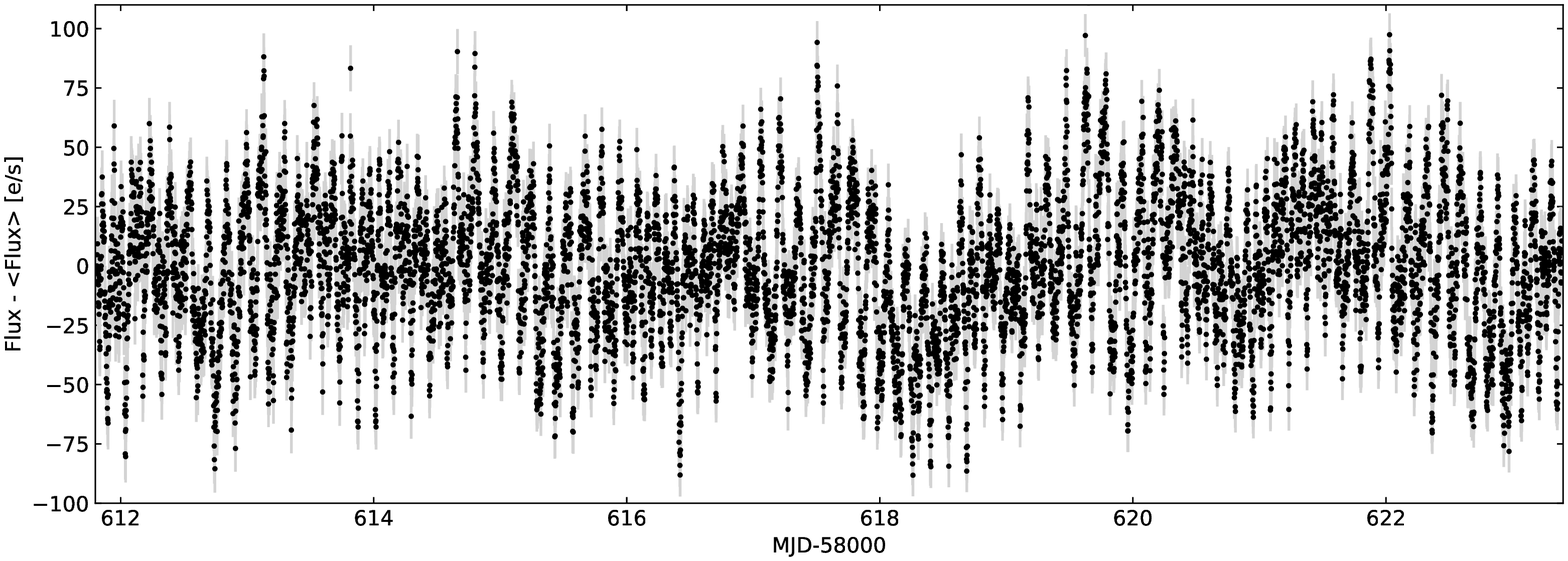}}
\resizebox{0.5\hsize}{!}{   \includegraphics{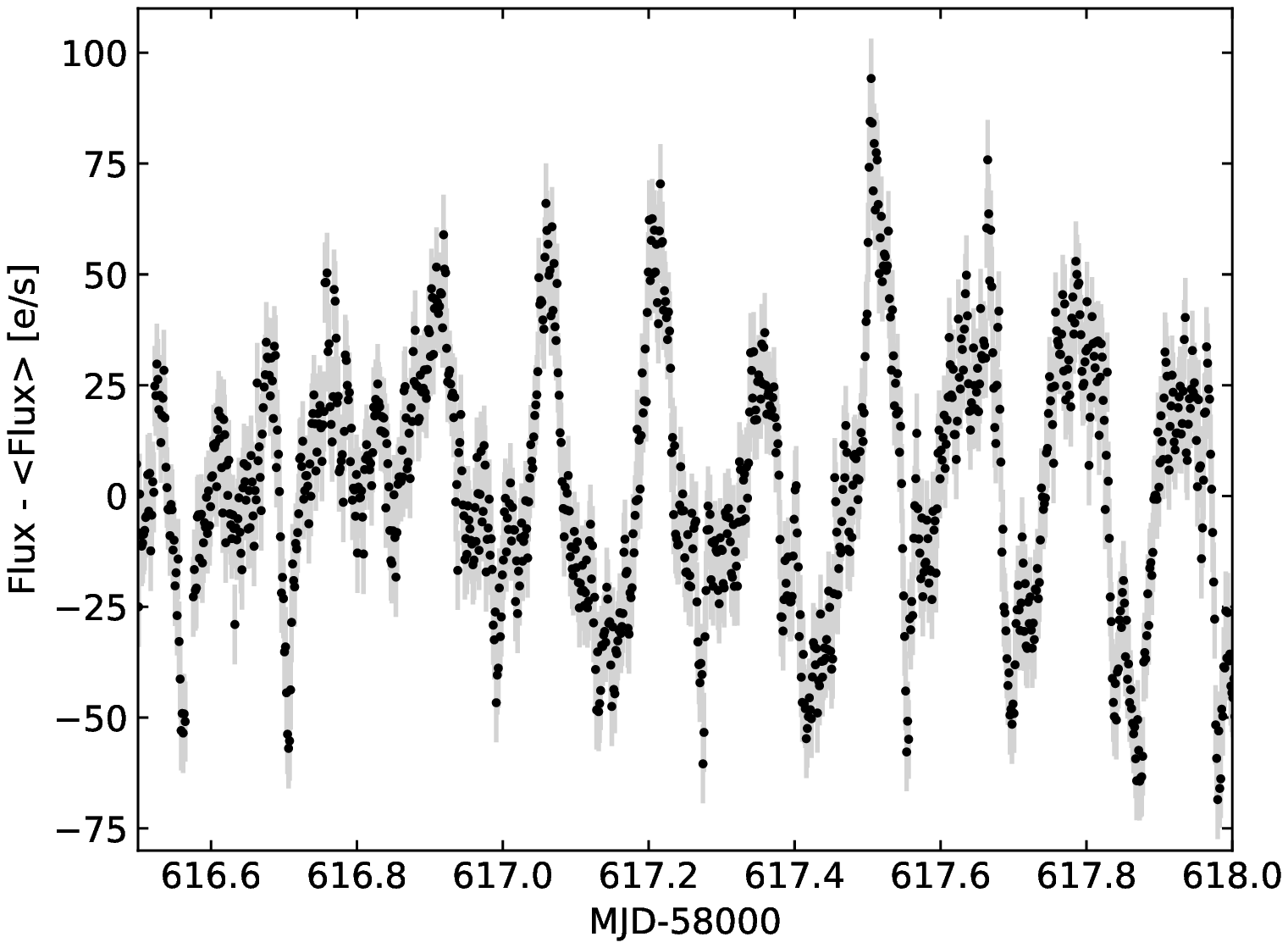}}\resizebox{0.5\hsize}{!}{   \includegraphics{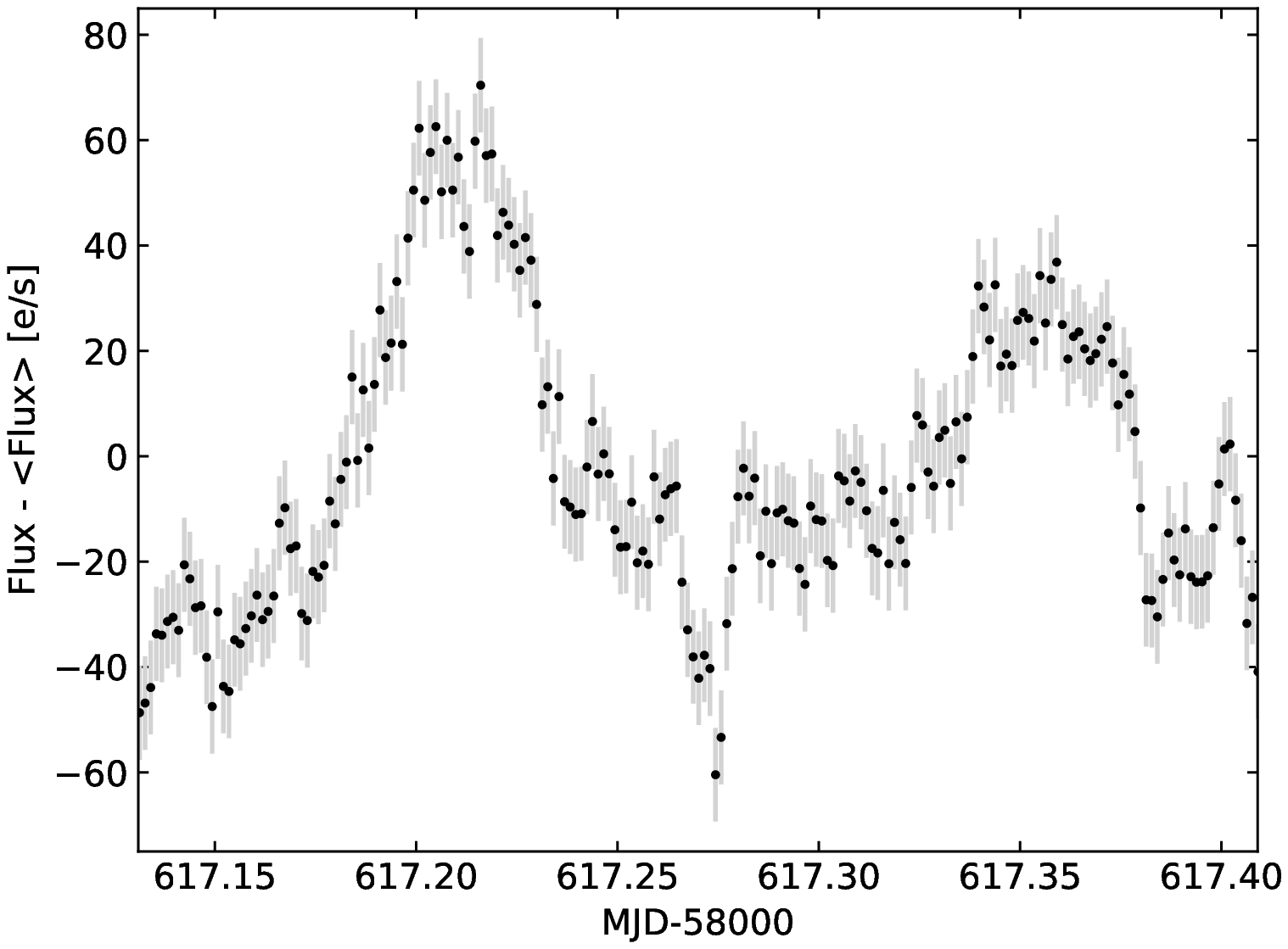}}
\caption{Representative samples of the TESS light-curve of AQ~Men (black points) with corresponding errors (grey lines) at different timescales. The data is from \textit{TESS} sector 11.}
\label{fig:example_lc}
\end{figure*}

AQ~Men was observed by the Transiting Exoplanet Survey Satellite (\textit{TESS}; \citealt{2015JATIS...1a4003R}) with 120\,s cadence during sectors 1, 5, 8, 11, 12, and 13. This corresponds to non-continuous observations in the time period between MJDs 58324 and 58681. Each sector lasts approximately 28~days. The start and end dates of individual sectors are available on the \textit{TESS} website\footnote{https://tess.mit.edu/observations/}. We used data reduced with the SPOC pipeline \citep{2016SPIE.9913E..3EJ}. While the reduced data have had most of the spacecraft systematic trends removed, in order to compensate for any secular changes in the telescope between the sectors, we subtracted the mean flux from data obtained during each of the sectors.  A sample of the \textit{TESS} light-curve is presented in Fig.~\ref{fig:example_lc}.  The full light-curve is available at the Mikulski Archive for Space Telescope (MAST\footnote{http://archive.stsci.edu/}).

\begin{figure}
\resizebox{\hsize}{!}{   \includegraphics{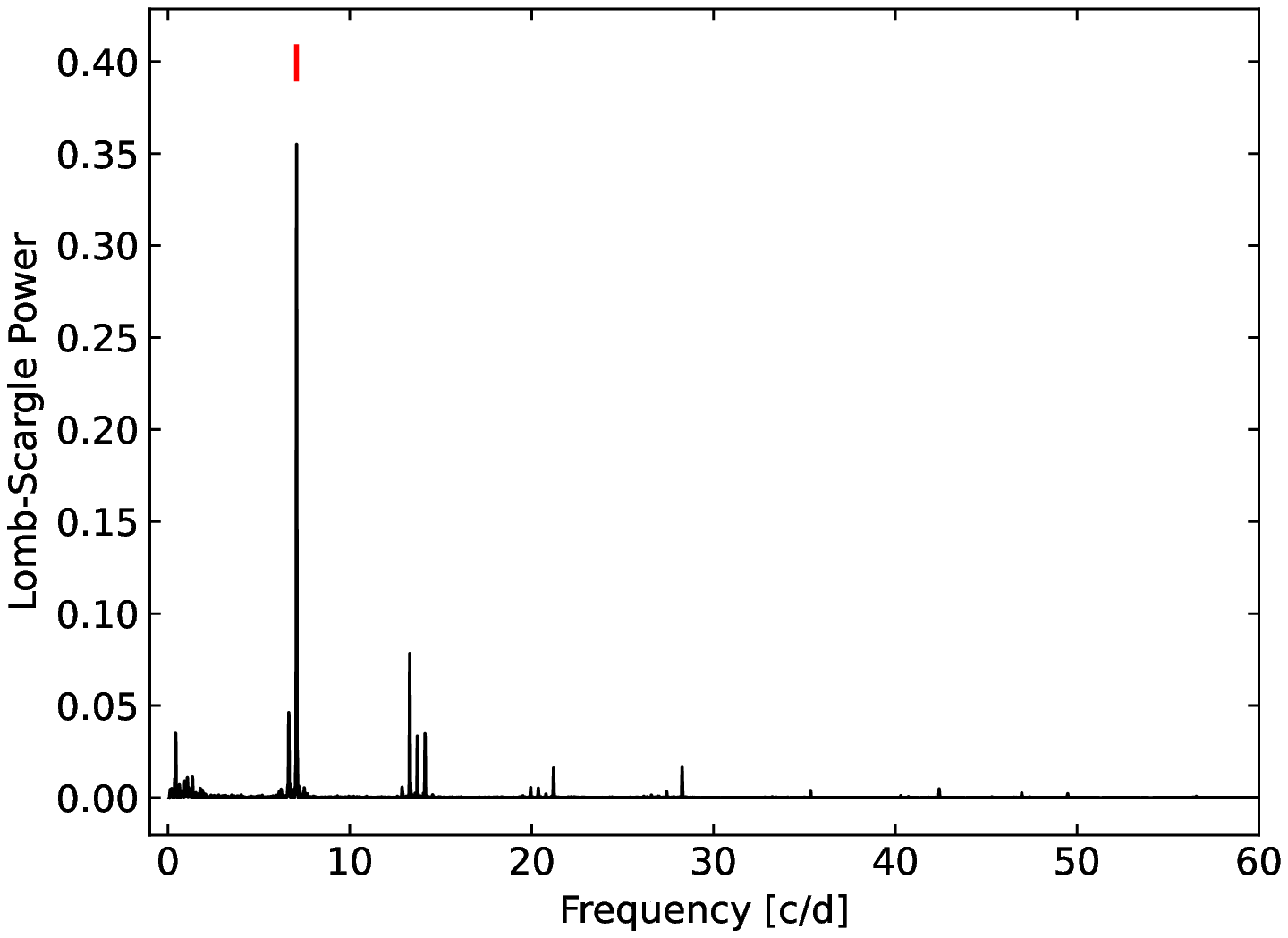}}
\resizebox{\hsize}{!}{   \includegraphics{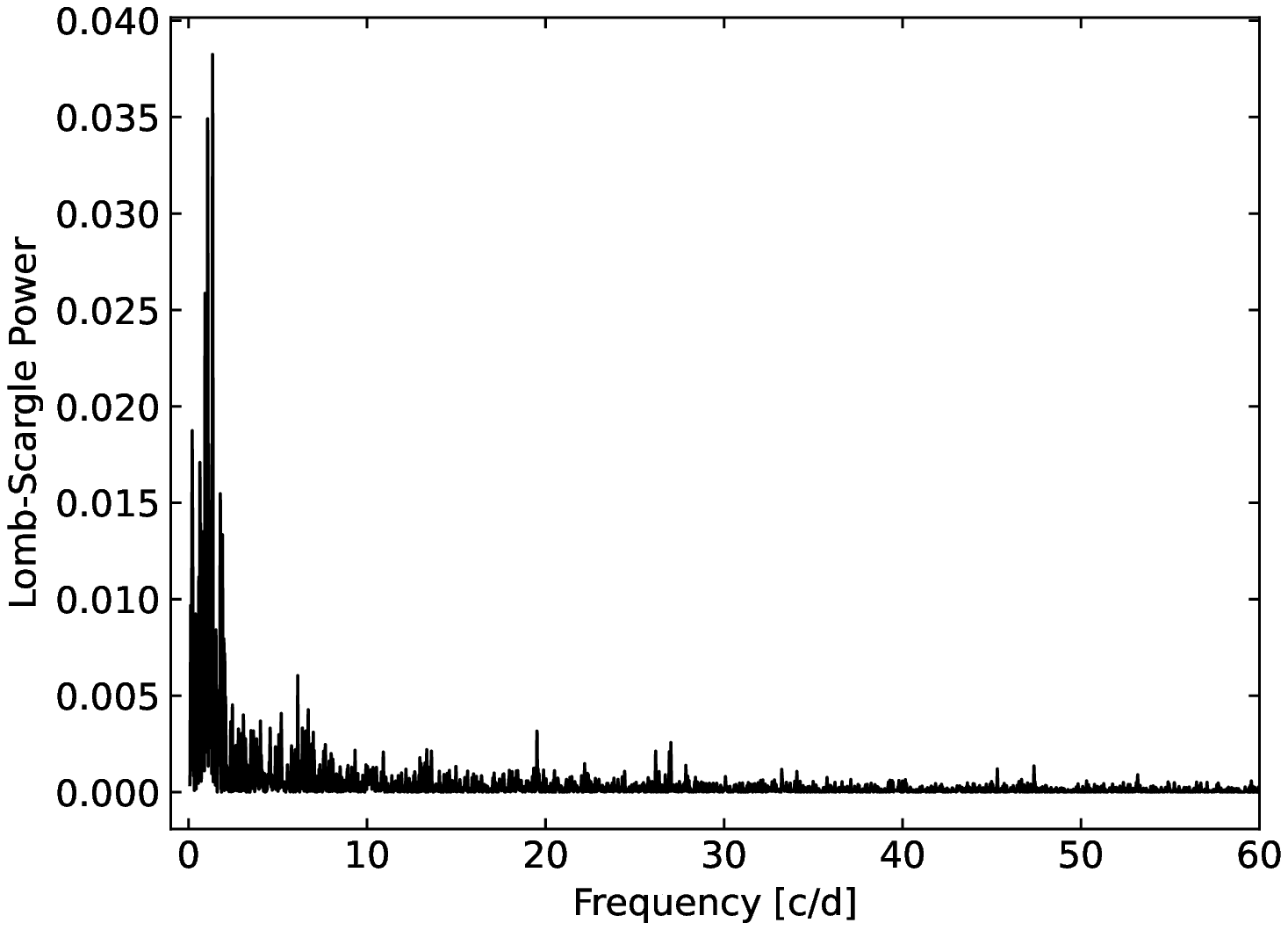}}
\caption{Top panel: A power spectrum of TESS AQ~Men observations taken from sector 13, where the orbital period is highlighted with a red line. Bottom panel: A power spectrum of the same data after undergoing the standard prewhitening procedure.}
\label{fig:powerspectrum}
\end{figure}

We searched for periodic signals in the data in individual sectors using a Lomb-Scargle periodogram \citep{1976Ap&SS..39..447L,1982ApJ...263..835S} that was calculated using a routine from Astropy \citep{astropy:2013, astropy:2018}. The power spectra were dominated by signals from the orbital period ($\omega_0$) and its harmonics (Fig.~\ref{fig:powerspectrum}). In order to search for weaker signals we used the standard consecutive prewhitening method \citep[e.g.][]{2019ApJ...878..155D}. We assumed that a signal is significant when its power exceeded 10 standard deviations of the nearby noise level. The power spectrum after prewhitening shows that there are no more significant signals remaining (Fig.~\ref{fig:powerspectrum}). The frequencies found in sector 13 data are presented in Table~\ref{tab:table_frq}. The results for other sectors were consistent with results from sector 13, with the exception that some signals are variable with time (Section~\ref{sec:evol}). Most notably, superorbital signal was not detected in sectors 5 and 8 due to the fact that the superorbital signal was weakest at those times.  We attempted the prewhitening method on the data from all \textit{TESS} sectors simultaneously but we were unable to remove certain signals, hinting that variability in AQ~Men can change its period or amplitude over time. The variability of the detected signals is discussed further in Section~\ref{sec:evol}.

Note that in the prewhitening process we removed the signals using the function
\begin{equation}\label{eq1}
 \mathrm{F(MJD)}=\sum_i \mathrm{A}_i*\sin [2\pi\omega_i(\mathrm{MJD}-\mathrm{MJD}_0)],
\end{equation}
therefore the MJD$_0$ values given in Table~\ref{tab:table_frq} correspond to the phase zero value of the sine in the $\mathrm{F(MJD)}$ function and not to the time of minimum value.  The mid-eclipse time in this notation is occurring at the orbital phase 0.6614.


\begin{table}
	\centering
	\caption{Frequencies of signals identified in AQ~Men observations from sector 13 of the \textit{TESS} together with their identification (ID): $\omega_0$ is the orbital frequency, N is the superorbital frequency, while $\omega_+$ and  $\omega_-$ are positive and negative superhump frequencies respectively. MJD$_0$ values corresponds to the phase argument of the sine function    described by Eq.~\ref{eq1}.}
	\label{tab:table_frq}
	\begin{tabular}{crrl} 
		\hline

ID &	Frequency [c/d]	&	Amplitude [e/s]	&	MJD$_0$		\\
		\hline
$\omega_0$      &	7.06869(13)	    &	24.66(17)	&	58667.68379(16)	\\
2$\omega_+$  &	13.29413(28)	&	11.25(17)	&	58667.67647(18)	\\
$\omega_+$      &	6.64591(34)	    &	9.42(17)	&	58667.71802(42)		\\
N               &	0.42093(41)	    &	7.78(17)	&	58668.449(8)	\\
2$\omega_0$     &	14.13860(41)	&	7.70(17)	&	58667.64725(24)	\\
2$\omega_0$-N   &	13.71513(44)	&	7.28(17)	&	58667.70072(27)	\\
4$\omega_0$     &	28.27526(60)	&	5.29(17)	&	58667.68000(18)		\\
3$\omega_0$     &	21.20627(62)	&	5.10(17)	&	58667.69222(25)		\\
$\omega_-$      &	7.4890(10)	    &	3.42(17)	&	58667.69220(11)		\\
2$\omega_0$-3N  &	12.8742(11)		&	3.08(17)	&	58667.7019(7)		\\
3$\omega_+$  &	19.9415(11)		&	3.11(17)	&	58667.65870(43)		\\
3$\omega_0$-2N  &	20.3658(11)		&	2.91(17)	&	58667.69293(45)	\\
$\omega_0$-2N   &	6.2228(12)		&	2.84(17)	&	58667.7486(15)	\\
6$\omega_0$     &	42.4152(12)		&	2.71(17)	&	58667.66454(24)		\\
5$\omega_0$     &	35.3431(13)		&	2.56(17)	&	58667.67009(30)		\\
4$\omega_0$-2N  &	27.4305(14)		&	2.32(17)	&	58667.68465(42)		\\
7$\omega_0$-6N  &	46.9523(15)		&	2.11(17)	&	58667.68200(27)		\\
7$\omega_0$     &	49.4830(17)		&	1.89(17)	&	58667.68073(29)		\\
2$\omega_0$+N   &	14.5574(19)		&	1.68(17)	&	58667.6597(11)		\\
4$\omega_+$  &	26.5834(21)		&	1.53(17)	&	58667.68059(65)		\\
3$\omega_0$-N   &	20.7830(21)		&	1.51(17)	&	58667.67364(84)		\\
6$\omega_0$-5N  &	40.3054(24)		&	1.36(17)	&	58667.66312(49)		\\
8$\omega_0$     &	56.5535(29)		&	1.09(17)	&	58667.67454(44)		\\
6$\omega_0$-4N  &	40.7268(32)		&	0.99(17)	&	58667.66489(66)		\\
		\hline
	\end{tabular}
\end{table}

\section{Results}\label{sec:res}

The strongest signal detected by us was at 7.06869(13)\,c/d, which corresponds to the orbital frequency ($\omega_0$) detected by \citet{2013MNRAS.435..707A}. We did not detect signals at the  superorbital (N) and negative superhump ($\omega_-$) frequencies discovered by \citet{2013MNRAS.435..707A} at 0.263(3)\,c/d and 7.332(3), respectively.  However, we did detect two signals at 0.42093(41) and 7.4890(10)\,c/d (Table~\ref{tab:table_frq}). These signals are consistent with satisfying the relation $\omega_0+\mathrm{N}=\omega_-$, therefore we interpret them as superorbital and negative superhump periods respectively.  We did not detect the signal at 0.12\,c/d which was detected by \citet{2020NewA...7801369B} in observations taken at the same time as \textit{TESS} sector 2. We note that signals with similar frequencies to the \citet{2020NewA...7801369B} detection are present in some of the \textit{TESS} sectors; however, they have signal-to-noise ratios consistent with being low-frequency noise.

One of the signals detected by us is located at a frequency 6.64591(34)\,c/d, which we interpret as a positive superhump ($\omega_+$), as this signal is at a frequency that is a few percent lower than the orbital frequency. The difference $\omega_0-\omega_+=$0.42278(47)\,c/d. While this difference is relatively close to the superorbital frequency, it differs from the superorbital frequency by 6.3 times the sum of the standard deviations of $\omega_0-\omega_+$ and the superorbital frequency. Therefore, the superorbital frequency is only consistent with the $\omega_0+\mathrm{N}=\omega_-$ relation and is not associated with the positive superhump.

Moreover, we detected signals at seven harmonics of the orbital frequency, reflecting a non-sinusoidal shape of the orbital variability. Similarly, \citet{2013MNRAS.435..707A} detected five harmonics of the orbital frequency. The other detected frequencies are consistent with simple arithmetic combinations of the orbital and superorbital frequencies. Due to the fact that $\omega_0+\mathrm{N}=\omega_-$ and $\omega_0-\mathrm{N}\simeq\omega_+$, the identifications given in Table~\ref{tab:table_frq} are not unambiguous. For example, the frequency 7$\omega_0$-6N  could be identified as 6$\omega_+$+$\omega_0$. We decided to identify the additional signals as a combination of $\omega_0$ and N, with the exception of frequencies at multiple times the positive superhump frequency. However, we note that  in most cases of signal processing it is rare to see signals at frequencies such as 7$\omega_0$-6N, without detecting any signals at frequencies 7$\omega_0$-nN, where 0<n<6. Therefore, notations such as 6$\omega_+$+$\omega_0$ might be preferable. The signals at frequencies at multiple times the positive superhump frequency are due to the strongly non-sinusoidal shape of positive superhumps. It is worth noting that the signal at 2$\omega_+$ was significantly stronger than the signal at $\omega_+$.    A similar multitude of signals with frequencies at arithmetic combinations of the orbital and superorbital frequencies is sometimes observed in systems where both positive and negative superhumps are present \citep{1999PASP..111.1281S,2011ApJ...726...92Fs}.       Interestingly, even though we detected a total of 24 frequencies (Table~\ref{tab:table_frq}), we did not detect signals at frequencies $2\omega_- - \omega_+$ and $2\omega_- + \omega_+$, which were theoretically predicted for systems in which both negative and positive superhumps are present simultaneously \citep{2000ApJ...535L..39W}.

\begin{figure*}
\resizebox{\hsize}{!}{   \includegraphics{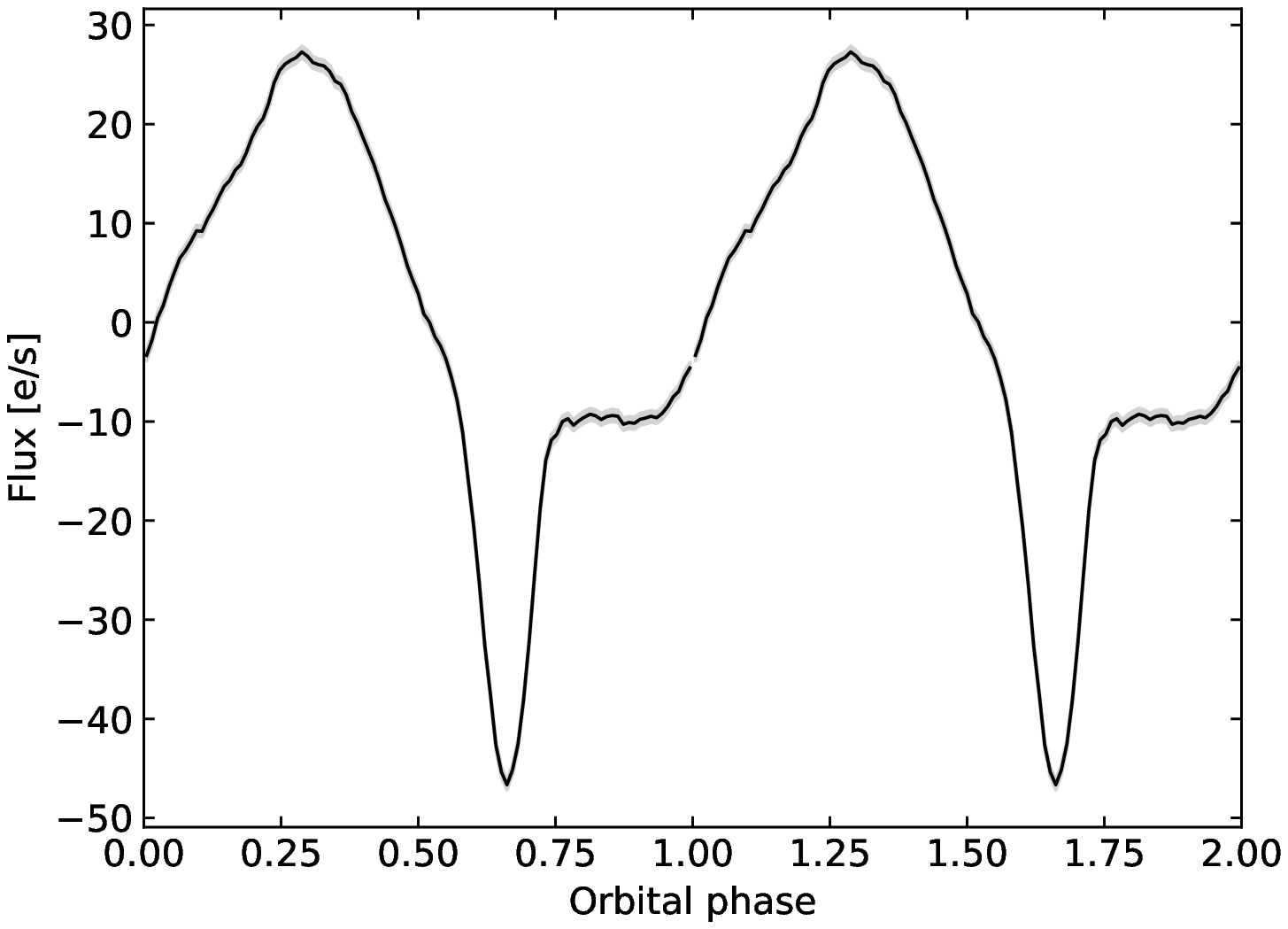} \includegraphics{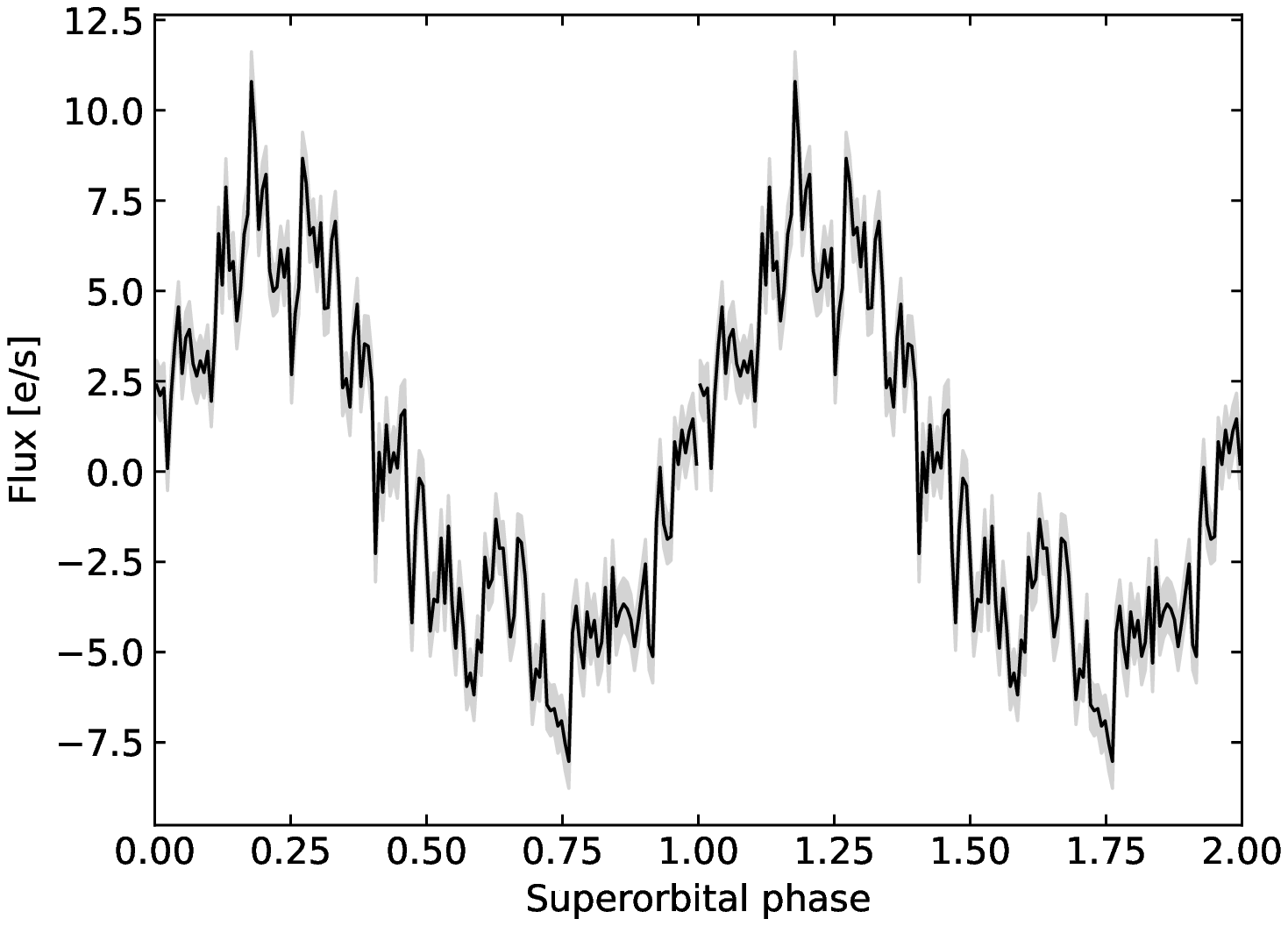}}
\resizebox{\hsize}{!}{   \includegraphics{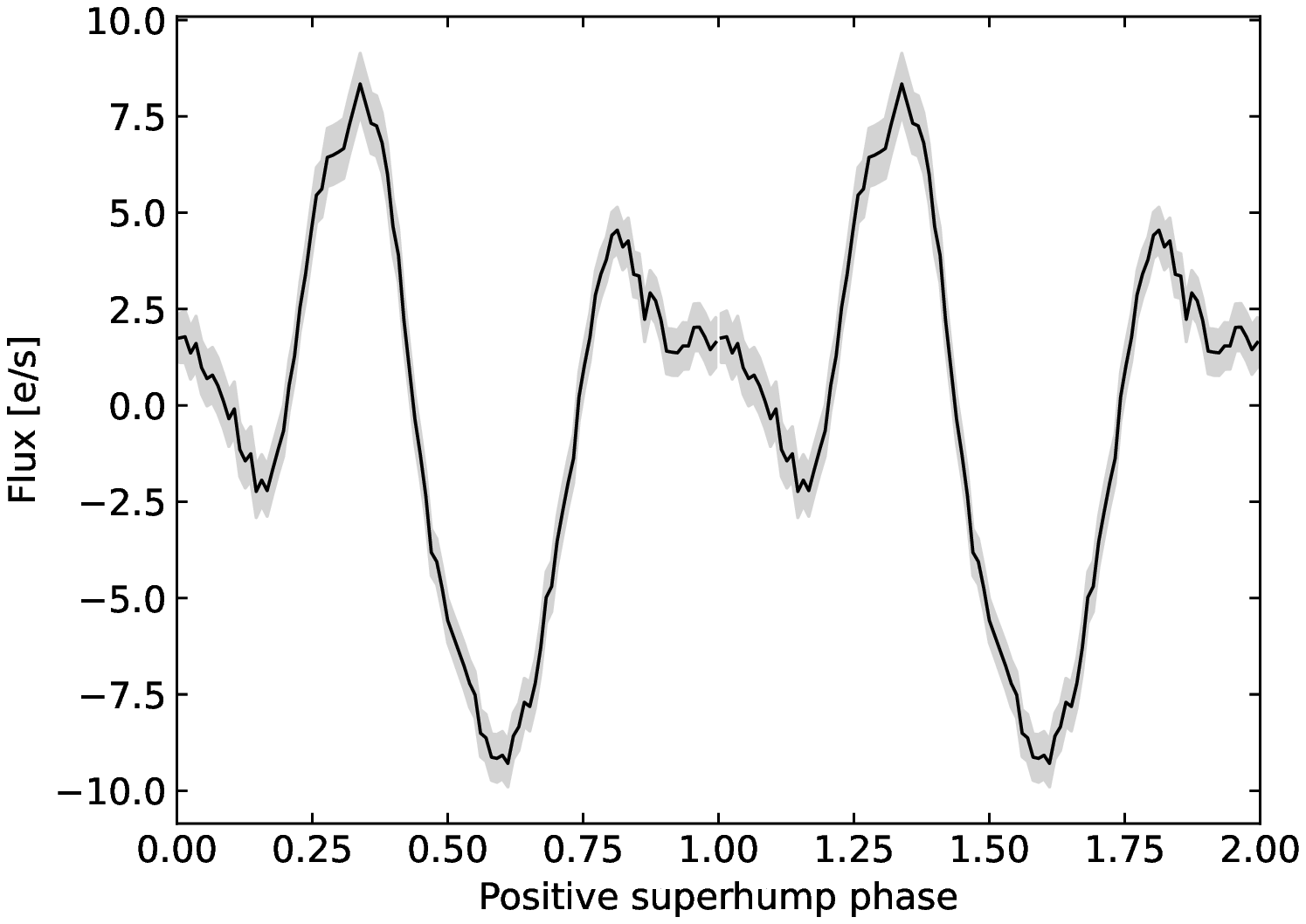} \includegraphics{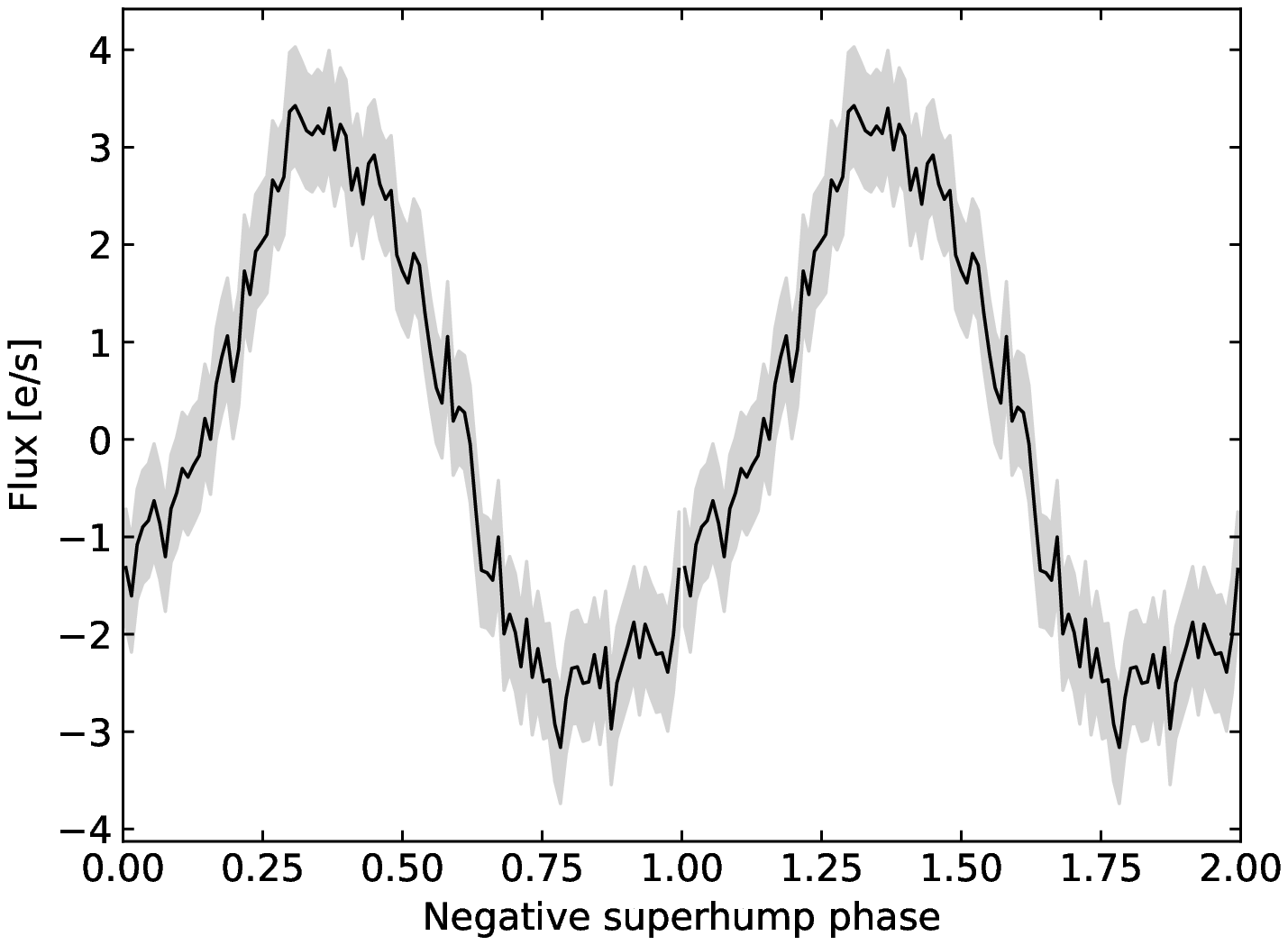}}
    \caption{The binned phase plots of orbital, superorbital, positive superhump and negative superhump variability (black lines). The grey area represents the standard error of the mean for each of the bins. The orbital phase is defined such that mid-eclipse occurs at orbital phase 0.6614.}
    \label{fig:binned_phased}
\end{figure*}

In order to study the shape of the orbital variability, we first prewhitened the data from all of the \textit{TESS} sectors. Afterward, we added back only the signals at integer multiples of the orbital frequency which were initially removed during the prewhitening. The same procedure was applied to the superorbital, negative superhump, and positive superhump signals together with their harmonics. The resultant light curves were then phased using the ephemerides from Table~\ref{tab:table_frq} and binned using 150 equally separated bins. The phase plots are presented in Fig.~\ref{fig:binned_phased}.

In the phase plot of orbital variability, an eclipse and an orbital hump are visible. Interestingly, the orbital hump appears before the eclipse, while in the observations of \citet{2013MNRAS.435..707A} the orbital hump maximum and the eclipse were aligned in time. The reason for the phase shift of the orbital hump is not clear. It has been accepted that orbital hump in systems such as AQ~Men is due to a presence of a hotspot. However, we note that an orbital hump due to a hotspot is rarely observed so long before the eclipse, and is rarely as prominent in nova-like variables as in the case of AQ~Men \citep[e.g.][]{1994AcA....44..257S,2013AcA....63..225R}. Since different explanations of an orbital hump have been proposed \citep[e.g. presence of spiral arms in the disc;][]{2015EAS....71..149K} the AQ~Men light curve needs to be modelled in detail in order to decisively confirm that the orbital hump is due to the hotspot.

The superorbital signal is sinusoidal, as is reflected by our detection of only the fundamental frequency. This is in agreement with the observations of \citet{2013MNRAS.435..707A}. The negative superhump is close to being sinusoidal while being slightly skewed and concave upward during the rise and concave downward during the decline. The drop from maximum is $\sim$20\% faster than the rise.  The opposite is true in the data of \citet{2013MNRAS.435..707A}, where the rise to the maximum is faster.

The shape of the positive superhump variability is strongly non-sinusoidal (Fig.~\ref{fig:binned_phased}). In fact, between phases $\sim$0.25 and $\sim$0.75 a pattern resembling a full cycle of a sinusoid appears, while between phases $\sim$0.75 and $\sim$1.25 there is a nearly linear drop in brightness. Positive superhumps are not expected to show such complex behaviour, as they typically show only small deviations from a sinusoid \citep[e.g.][]{2006A&A...456..599P,2011ApJ...741..105W,2013PASJ...65...76K,2020PASJ...72...49T}, even when a large number of signals are present in the periodogram \citep{1999PASP..111.1281S}. The exceptions are positive superhumps in dwarf novae, which can change their shape significantly during an outburst \citep[e.g.][]{1995PASP..107..657P}. The reason for a strongly non-sinusoidal shape of the positive superhump in AQ~Men is not clear.

\subsection{The superhumps and superorbital signal evolution}\label{sec:evol}

\begin{figure}
\resizebox{\hsize}{!}{   \includegraphics{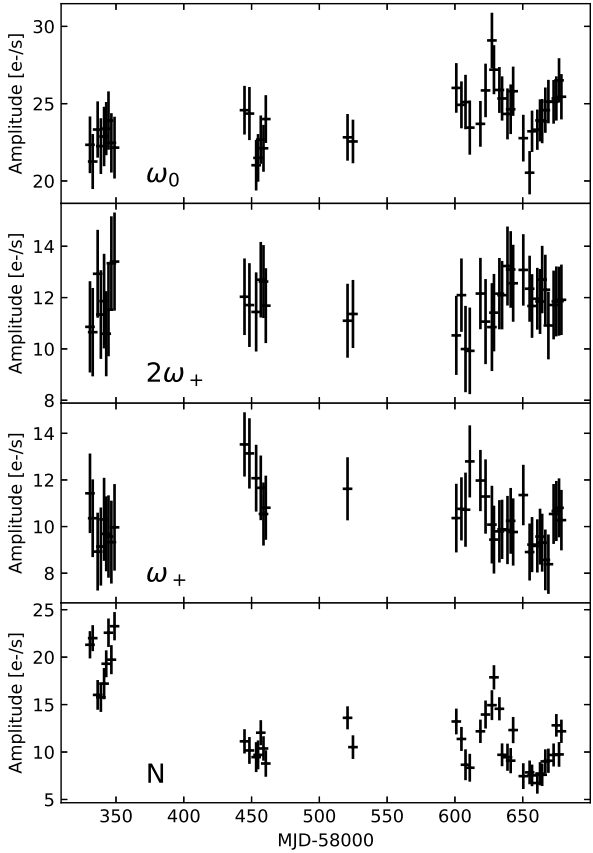}}
\caption{Plots showing the amplitude variability of the orbital, superorbital and positive superhump signals, as well as the amplitude of the first harmonic of the positive superhump signal. }
\label{fig:amplitudes}
\end{figure}

\citet{2013MNRAS.435..707A} did not detect a positive superhump in AQ~Men, while the negative superhump was present in their observations. In our data the positive superhump has a significantly greater amplitude than the negative superhump. This suggests that the amplitudes of these two forms of variability had to have changed significantly in the past. To explore this further, we calculated a moving dynamical periodogram of the \textit{TESS} data. The observations were divided in segments with lengths of 8 days, which roughly corresponds to three times the superorbital period.  The segments were taken every 2 days, meaning each data point was included in four separate segments. It is apparent that the superorbital signal is the second strongest signal in the earliest observations and at MJD$\sim$58630, while on some dates (e.g. MJD$\sim$58660) it is one of the weakest (Fig.~\ref{fig:amplitudes}). Moreover, the low-frequency noise sometimes had power similar to the signals from the system. Because of this, we performed a non-standard prewhitening method, where we removed the four strongest signals in the order given in Table~\ref{tab:table_frq}. Remaining signals were not strong enough to study using this method when such short segments were employed.

After MJD~58360 the amplitudes of orbital and superorbital variability appear to be correlated, with both of them having a maximum at MJD~58628 (Fig.~\ref{fig:amplitudes}). The correlation between the amplitudes of the superorbital and orbital periods is expected, as they both are related to the brightness of the accretion disc. Between MJD~58400 and MJD~58550 the amplitude of neither the orbital nor superorbital variability changes significantly. Before MJD~58360 the amplitude of the superorbital signal was at its highest within the period covered by \textit{TESS} observations, but the amplitude of the orbital signal was the same as at MJD~58400--58550. This is unexpected, but may be explained if the accretion disc is not entirely eclipsed. This occurs when the system is slightly inclined, and the donor eclipses only a part of an out of plane accretion disc (grazing eclipses; \citealt{2013MNRAS.435..707A}). In this scenario only the central parts of the accretion disc were brighter before MJD~58360 than during the later observations. Since the central parts of the accretion disc would not not be eclipsed, a change in their brightness would not change the eclipse depth. However, these parts of the accretion disc are still precessing and their brightness would reflect on amplitude of the superorbital signal.

\begin{figure}
\resizebox{\hsize}{!}{   \includegraphics{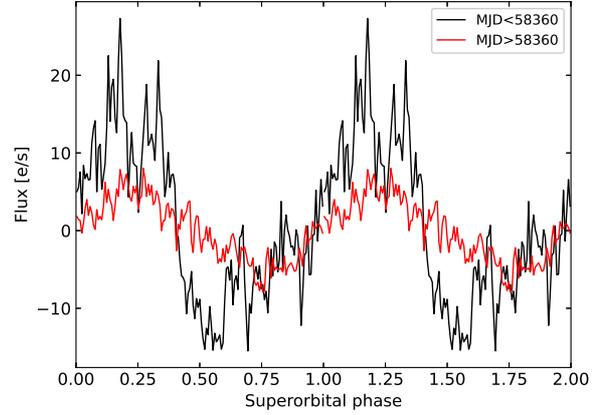}}
\caption{Same as Fig.~\ref{fig:binned_phased}, but only for superorbital variability and with data divided at MJD~58600 (see text).}
\label{fig:mean_superorb_mjd}
\end{figure}

In order to check whether the shape of superorbital variability changed with the system brightness, we plotted the mean superorbital signal obtained in the same way as for Fig.~\ref{fig:binned_phased}, but with data separated before and after MJD~58360. The result is presented in Fig.~\ref{fig:mean_superorb_mjd}. The shape of the superorbital signal changes significantly between the two periods of time. At MJD>58360 the shape is sinusoidal, as is expected for variability caused by a precessing tilted accretion disc. At MJD<58600 the shape is a skewed sinusoid and the rise is twice as long as the decline. The reason for this change of the shape of the superorbital variability is not known.

The superhump periods in some nova-like variables seem to be stable over several years \citep{2012NewA...17..433R}. However, variability of the superhump frequency is observed in dwarf novae during superoutburst \citep[e.g.][]{2019MNRAS.488.4149C,2020PASJ...72...14K} and is sometimes observed in nova-like variables \citep[see e.g.][and references therein]{2003ASPC..292..313A}. The negative superhump and superorbital frequencies in AQ~Men have clearly changed in the past, given that the frequencies measured by \citet{2013MNRAS.435..707A} and by us differ significantly. The discovery of the superorbital period at 0.263(3)\,c/d was based on data from MJD$\sim$52333 \citep{2013MNRAS.435..707A}. The mean date of all of the \textit{TESS} data points is MJD~58541, which gives a difference of 6208\,d. The frequency change of the superorbital signal between those two observations was 0.158(4)\,c/d.

\begin{figure}
\resizebox{\hsize}{!}{   \includegraphics{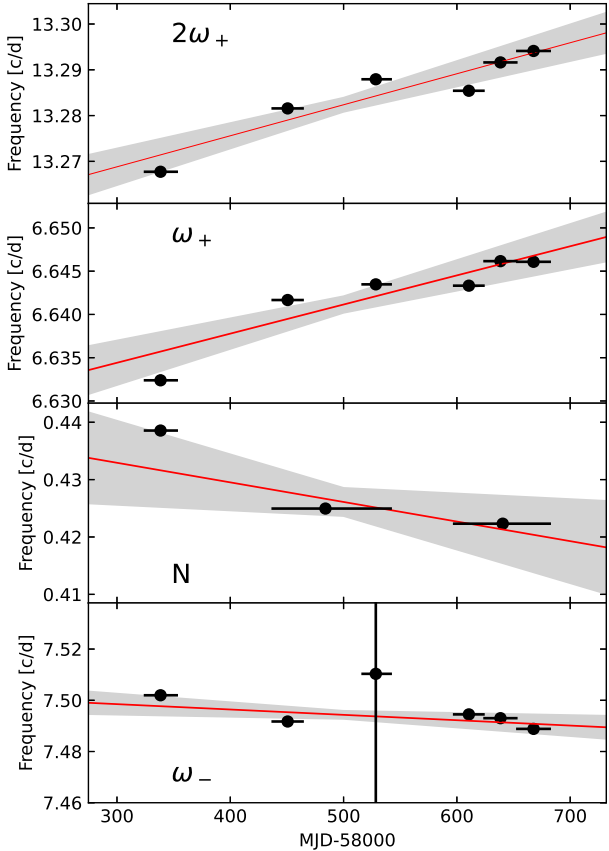}}
\caption{Plots showing the frequency variability of the superorbital, positive superhump and negative superhump signals, as well as the frequency of the first harmonic of the positive superhump signal. The fit to the points is plotted with a red line. Gray areas represents the error of each fit.}
\label{fig:frequencies}
\end{figure}

In order to measure the possible changes in the detected frequencies, we used the standard prewhitening method for each sector. The only exception was analysis of the superorbital signal, for which we divided the \textit{TESS} data into three segments. The first segment consisted of sector 1, the second segment of sectors 5 and 8, and the third segment of sectors 11, 12, and 13. This was done because the superorbital signal had signal to noise ratio too low for some of the sectors to measure the frequency change with sufficient accuracy (Fig.~\ref{fig:amplitudes}). In order to estimate the rate of frequency change of the signals, we assumed that they changed linearly and we fitted lines to each of them individually. As a result, we estimated the upper limit of the rate of change of the orbital frequency $|\dot{\omega_0}|<2.12 \times 10^{-6}$\,c/d$^{2}$. The apparent constant orbital period is expected for the relatively short time span covered by \textit{TESS} observations. In particular, for the AQ~Men orbital period of 3.4\,h, the expected $\dot{\omega_0}$ is of order of $10^{-11}$\,c/d$^{2}$ (see e.g. fig.~11 of \citealt{2011ApJS..194...28K}) which is well below our accuracy. In the case of the signal at frequency 2$\omega_+$, the estimated rate of change is $\dot{(2\omega_+)}= (6.8\pm 1.3)\times 10^{-5}$\,c/d$^{2}$, showing a clear increase of frequency with time. Similarly, the rate of change we measure for the positive superhump is $\dot{\omega}_+= (3.4\pm 0.8)\times 10^{-5}$\,c/d$^{2}$, which is consistent with half of the rate we measure for the signal at $2\omega_+$. This rate is an order of magnitude higher than the rates observed in dwarf novae \citep[e.g.][]{2017PASJ...69...75K}, but is not unexpected for a nova-like variable \citep[e.g.][]{2013PASJ...65...76K}.      In the case of the superorbital frequency, we estimated $\dot{\mathrm{N}}= (-3.4\pm 2.4)\times 10^{-5}$\,c/d$^{2}$. In the case of the negative superhump frequency we measured $\dot{\omega}_-= (-2.1\pm 1.3)\times 10^{-5}$\,c/d$^{2}$. The rates of change of the negative superhump and superorbital frequencies are equal within the errors, which is expected given that they follow the equation $\mathrm{N}=\omega_--\omega_0$. However, we note that the errors of $\dot{\mathrm{N}}$ and $\dot{\omega}_-$ are relatively large and relation between them should be accepted with care. Our fits are presented in Fig.~\ref{fig:frequencies}.

The rate of superorbital frequency change $\dot{\mathrm{N}}= (-3.4\pm 2.4)\times 10^{-5}$\,c/d$^{2}$ means that the rate of change the of superorbital frequency has inverted at least once in the past and is now decreasing back towards the frequency measured by \citet{2013MNRAS.435..707A}.  The superorbital frequency and, by extension, the negative superhump frequency, can change with system brightness \citep{2013PASJ...65...76K}. It is not possible to determine whether this is the reason in the case of AQ~Men on a longer timescale, as our data and the data of \citet{2013MNRAS.435..707A} and  \citet{2020NewA...7801369B} were taken with different filters. However, during the period covered by \textit{TESS} alone, AQ~Men was fading nearly linearly with a rate of $-1.9\pm 0.3$~mJy/d (Fig.~\ref{fig:ASASSN}). This is consistent with the correlation between the system brightness and superorbital frequency.

Interestingly, the measured rate of positive superhump frequency change is consistent with the negative value of the rate measured for the superorbital and negative superhump signals. Correlated changes of negative and positive superhump frequencies have been observed in the past and they were similarly connected to changes in brightness of the system \citep{2013PASJ...65...76K}. A longer time span of data is necessary to confirm the relation between the frequency changes of the positive superhump and the superorbital signal.

Since the superorbital frequency is changing towards the one observed by \citet{2013MNRAS.435..707A}, one can hypothesize that the positive superhump will fade away and only the negative superhump will be visible, similar to what was seen in the observations of  \citet{2013MNRAS.435..707A}. One may suspect that when the superorbital frequency in a given system would be the same, a similar accretion disc state can be presumed. Such switches between positive and negative superhump states have been observed in another nova-like variable, TT~Ari \citep[e.g.][]{1975AcA....25..379S,1998ApJ...503L..67S,1999A&A...351..607K,2009A&A...496..765K,2009CoAst.159..114W}. Moreover, on rare occasions, positive and negative superhumps in TT~Ari have been detected simultaneously \citep{2013AstL...39..111B}, further expanding similarities between TT~Ari and AQ~Men.

\subsection{Exploring the precession of the accretion disc}\label{sec:supvar}

\begin{figure}
\resizebox{\hsize}{!}{   \includegraphics{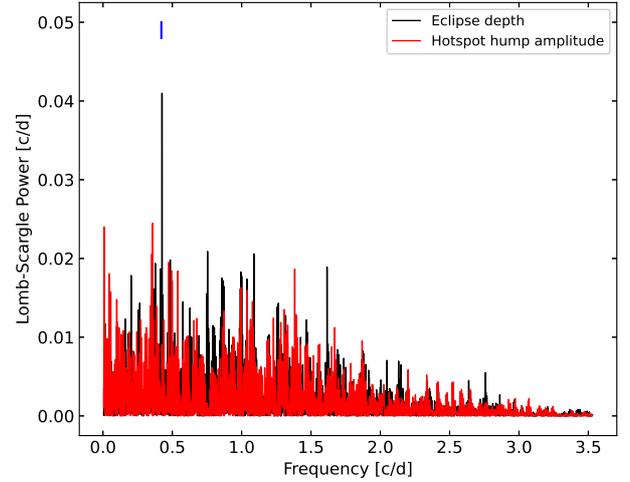}}
\caption{Lomb-Scargle power spectra of eclipse depths and hotspot hump height. The superorbital frequency detected in the measurements of the eclipse depths is highlighted with a blue line. }
\label{fig:Amplitudes_powerspec}
\end{figure}

\citet{2013MNRAS.435..707A} discovered that the eclipses in AQ~Men change their depth significantly between orbital cycles. It was discussed by them that this may be an effect of grazing eclipses which change their depth depending on the phase of superorbital variability, i.e. eclipses happening at different accretion disc orientations with respect to the observer. However, their data were not sufficient to confirm this hypothesis. In order to explore the variability in the depths of the eclipses, we divided the data into segments with lengths equal to twice the orbital period and segments separated by one orbital period length. We then folded the data using the orbital solution given in Table~\ref{tab:table_frq}. Using phased data, we measured the minimum flux during each eclipse using the mean flux between phases 0.657--0.666, the baseline flux using the mean flux between phases 0.8--0.9, and the hotspot hump maximum using the mean flux between phases 0.27--0.33 (Fig.~\ref{fig:binned_phased}). We then calculated the depth of each eclipse by subtracting the eclipse minimum from the quiescence flux, and the hotspot hump height by subtracting the quiescence flux from the hump maximum. Each measurement was assigned the mean time of each segment. For this dataset, we calculated a Lomb-Scargle periodogram (Fig.~\ref{fig:Amplitudes_powerspec}). The hotspot hump height did not show any periodic changes. Meanwhile, the eclipse depth clearly varied with the superorbital period, which confirms the hypothesis of \citet{2013MNRAS.435..707A}. 

\begin{figure}
\resizebox{\hsize}{!}{   \includegraphics{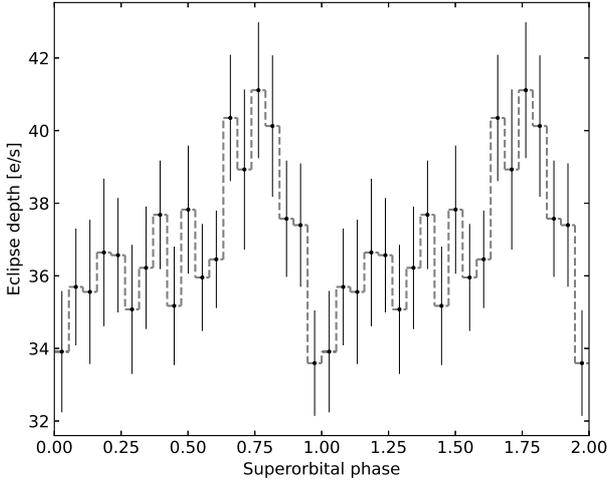}}
\caption{Eclipse depth as a function of superorbital phase. The errors represent the standard error of the mean.}
\label{fig:Eclipse_depth_phased}
\end{figure}

In order to study how the eclipse depths change with the superorbital period, we phased the eclipse depths with the superorbital signal solution from Table~\ref{tab:table_frq}. The eclipse depths rose slowly from superorbital phase $\sim$0.0 to $\sim$0.75, and decreased $\sim$3 times as fast (Fig.~\ref{fig:Eclipse_depth_phased}). Interestingly, the maximum eclipse depth at superorbital phase $\sim$0.75 appears to correspond to the minimum of the superorbital variability (Fig.~\ref{fig:binned_phased}). 

\begin{figure}
\resizebox{\hsize}{!}{   \includegraphics{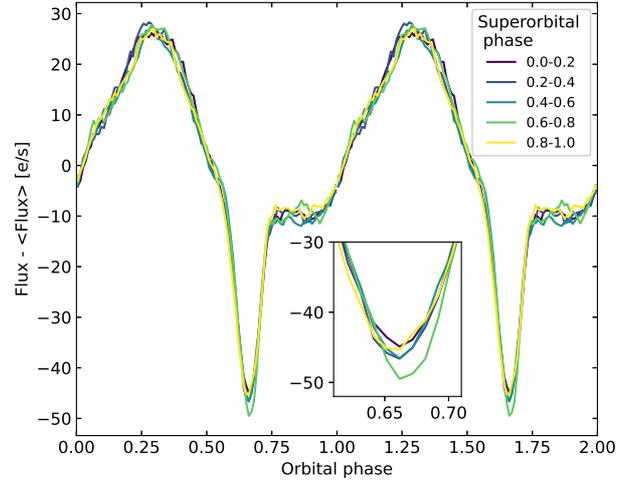}}
\resizebox{\hsize}{!}{   \includegraphics{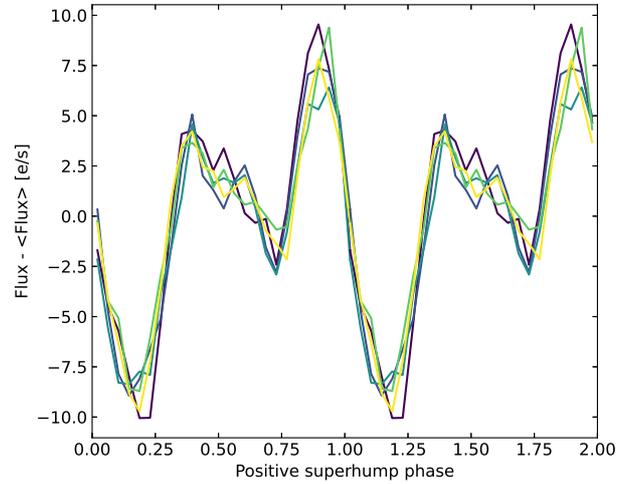}}
\resizebox{\hsize}{!}{   \includegraphics{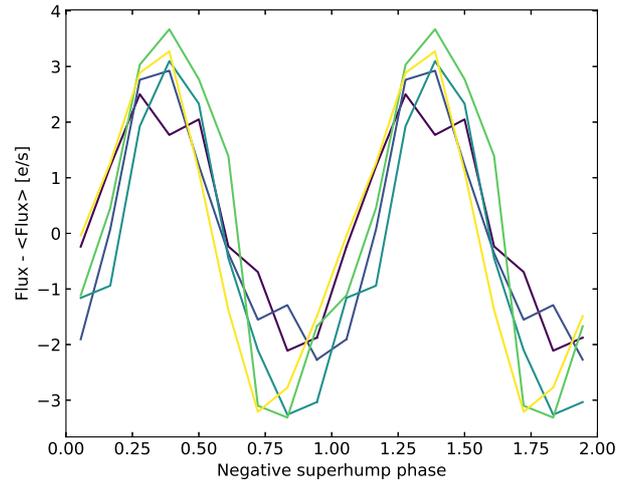}}
\caption{Same as Fig.~\ref{fig:binned_phased}, but with data points separated at different superorbital variability phases. The orbital modulation has been plotted using 100 bins, the positive superhump using 25 bins, and negative superhump using 10 bins. }
\label{fig:Orb_change_superorb}
\end{figure}

To explore the possibility of any other changes in the shape of the variability, we reproduced Fig.~\ref{fig:binned_phased} but separated data points by the associated phases of the superorbital variability. In other words, we studied how the orbital variability, positive superhump, and negative superhump behave at different phases of the superorbital variability. The resultant phase plots are presented in Fig.~\ref{fig:Orb_change_superorb}. It is clear that when it comes to the shape of the orbital modulation, only the eclipse depth is changing with the superorbital phase. The amplitude of the positive superhump seems to vary slightly with the superorbital phase, but no obvious correlation between the superorbital phase and the positive superhump shape is apparent. The amplitude of the negative superhump is too low to study its shape reliably using this method. However, it seems that the negative superhump has a maximum amplitude at superorbital phase $\sim$0.7 and a minimum amplitude at superorbital phase $\sim$0.1, similarly to the depth of the eclipses.

 \citet{2013MNRAS.435..707A} argued that the changes of superhump and orbital variability shapes with superorbital phase can serve as a test of accretion disc geometry. In particular, \citet{2012ApJ...745L..25M} showed that the accretion stream might switch from under to over the tilted accretion disc every half of the disc precession period, creating a lift that can push the disc out of the orbital plane. \citet{2013MNRAS.435..707A} hypothesised that a study of the shape of the negative superhump and eclipse depths can serve as a test of this model. The comparison of the models of \citet{2012ApJ...745L..25M} to our data is not trivial, as \citet{2012ApJ...745L..25M} did not simulate light curves of the variability. The fact that we observe the maximum amplitudes at superorbital phase $\sim$0.75 and minimum at phase $\sim$0 seems to not be in agreement with the simulations of \citet{2012ApJ...745L..25M}, as in there the switch between under- and over-disc accretion happens at every half of the superorbital phase, and one would expect the minimum and maximum to be separated by half of the superorbital period. However, \citet{2012ApJ...745L..25M} simulated only one case, with an accretion disc being out of the orbital plane by 5 degrees. Therefore, our observations may be consistent with this model if different disc inclinations or disc sizes are considered.

\section{Discussion}\label{sec:discussion}

A common method used to describe superhump periods is a positive superhump period excess $\epsilon_+=(P_{+}-P_{0})/P_{0}$ and a negative superhump period deficit $\epsilon_-=(P_{0}-P_{-})/P_{0}$, where $P_{0}$ is the orbital period, $P_{+}$ is the positive superhump period and $P_{-}$ is the negative superhump period. Using the periods from Table~\ref{tab:table_frq}, we obtain $\epsilon_-=0.056$ and $\epsilon_+=0.064$. \citet{2013MNRAS.435..707A} observed only the negative superhump in AQ~Men, and for their observations from the year 2002 we obtain $\epsilon_-=0.036$. 

\begin{figure}
\resizebox{\hsize}{!}{   \includegraphics{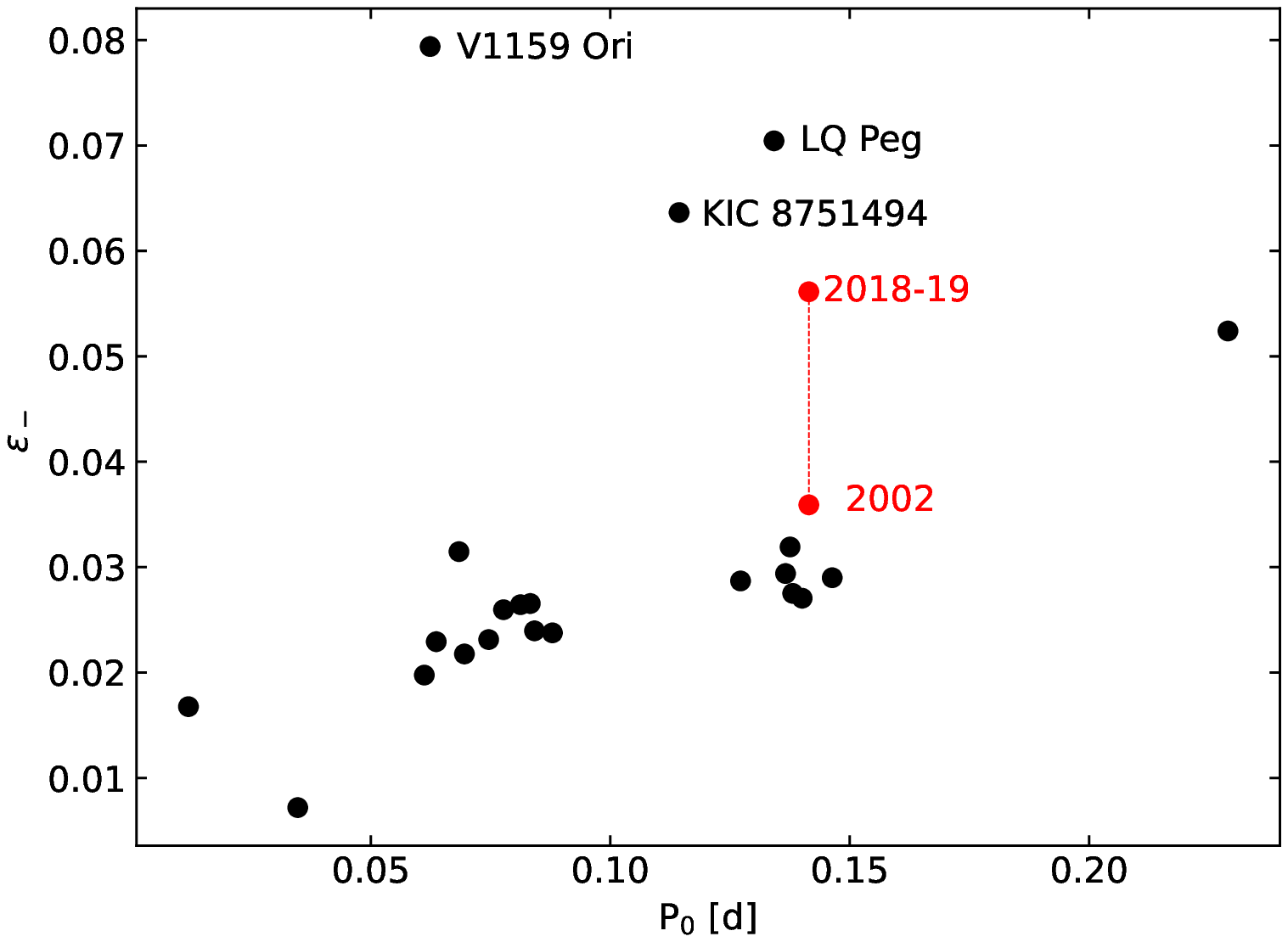}}
\resizebox{\hsize}{!}{   \includegraphics{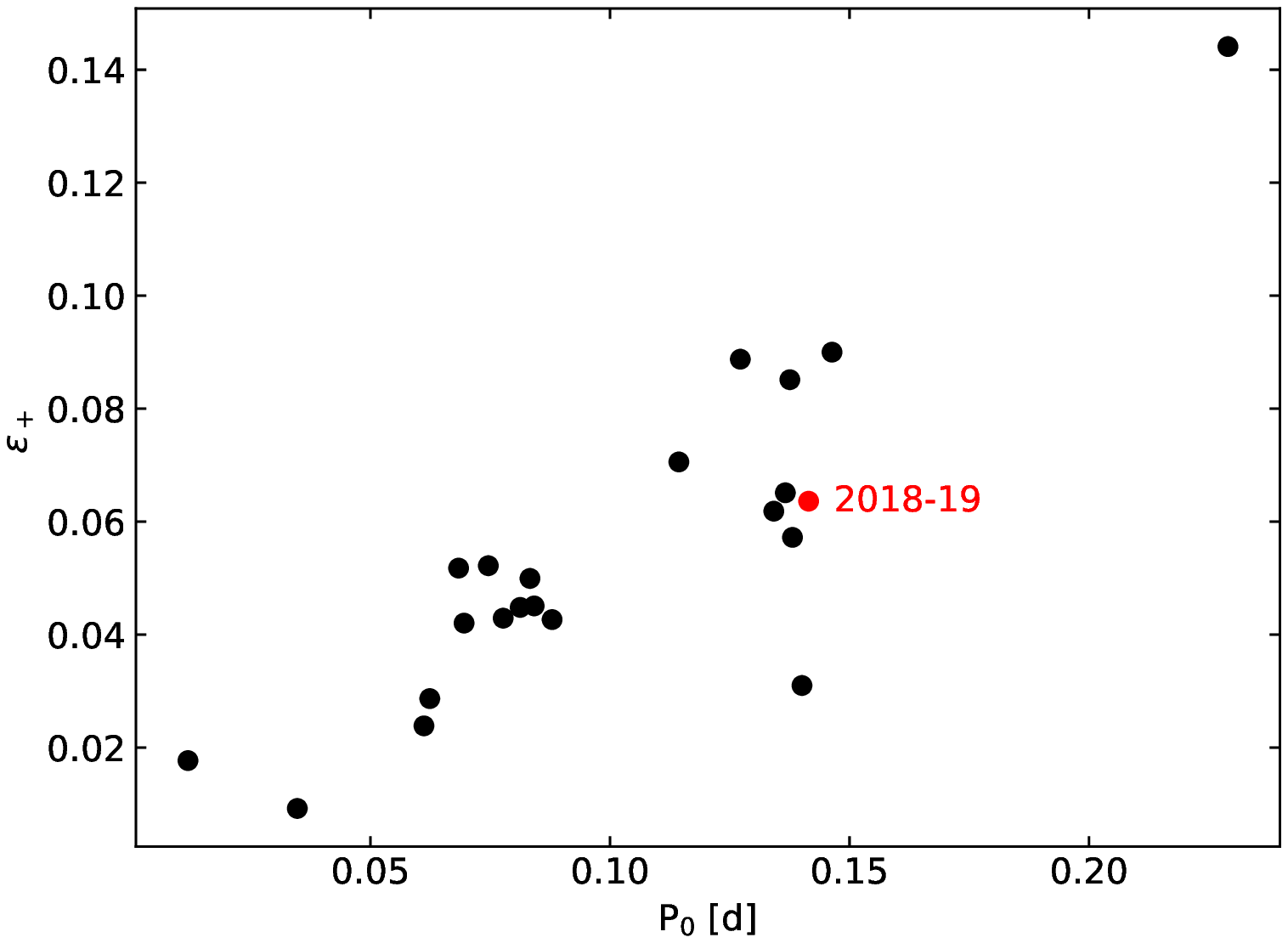}}

\caption{Relations between the superhump excess/deficit and the orbital period of a number of cataclysmic variables. The AQ~Men values (red points) from the year 2002 are from \citet{2013MNRAS.435..707A} and the values from 2018-2019 are from this work. The black points are previously known systems \citep[][and references therein]{2014PASJ...66...67O}.}
\label{fig:Porb_eps}
\end{figure}

The superhump excess/deficit has been shown to be correlated with the orbital period of the system \citep[e.g.][]{1999dicb.conf...61P,10.1046/j.1365-8711.2002.05280.x} and, consequently, with the mass ratio of the system components \citep[e.g.][]{2001MNRAS.325..761M,2005PASP..117.1204P}. The value of the period deficit observed in 2002 in AQ~Men was close to the typical value of $\epsilon_-\simeq 0.03$ that is expected for its orbital period (Fig.~\ref{fig:Porb_eps}). During the \textit{TESS} observations, the period deficit value was closer to the other outliers in this parameter space: V1159~Ori, LQ~Peg, and KIC~8751494. This could be interpreted as being due to the fact that the accretion disc in AQ~Men was in a state that is rarely observed in cataclysmic variables. Similarly to AQ~Men, LQ~Peg and KIC~8751494 are also nova-like systems, while V1159~Ori is a dwarf nova. On the other hand, the period excess observed by \textit{TESS} is consistent with the values for other cataclysmic variables at the same orbital period (Fig.~\ref{fig:Porb_eps}).

\begin{figure}
\resizebox{\hsize}{!}{   \includegraphics{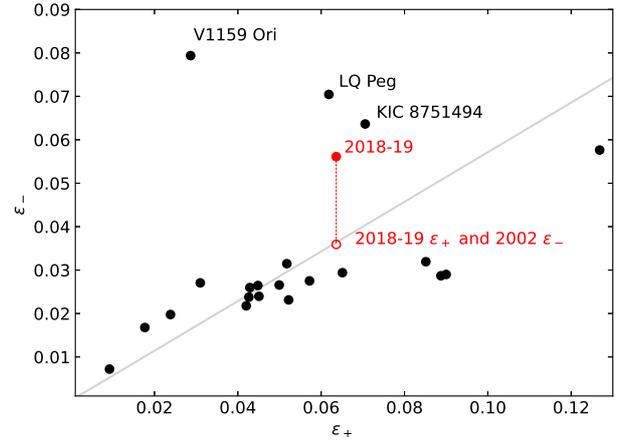}}
\caption{Same as Fig.~\ref{fig:Porb_eps}, but for the relation between period excess and period deficit. The theoretically predicted 7/4 ratio is shown as grey line.}
\label{fig:eps_pm}
\end{figure}

In a system where both positive and negative superhumps are present, the ratio of period excess to period deficit has been shown to be usually equal to $\sim$2 \citep{1997PASP..109..468P,2009MNRAS.398.2110W}. When pressure is not affecting the accretion disc, this ratio was theoretically shown to be close to 7/4 \citep{1998MNRAS.299L..32L,2009ApJ...705..603M,2013PASJ...65...95O}.  In the case of AQ~Men, the ratio of period excess to period deficit in the \textit{TESS} observations differ significantly from the main trend (Fig~\ref{fig:eps_pm}). In the case of KIC~8751494 the deviation from the 7/4 ratio can be explained by pressure effects \citep{2013PASJ...65...76K}, which can also explain the ratio observed in AQ~Men. In the case of V1159~Ori, the deviation from the 7/4 ratio can be explained by impulsive negative superhumps that were proposed by \citet{2013PASJ...65...95O}. While the nature of impulsive negative superhumps is not known, they differ from regular negative superhumps in that they occur only during a decline from a dwarf nova outburst, they are observed only for a few days and their frequency is highly variable \citep{2011ApJ...741..105W,2013PASJ...65...95O}. The impulsive negative superhump interpretation is unlikely in the case of AQ~Men, as no dwarf nova outburst was observed and the negative superhump frequency remained relatively stable during all the \textit{TESS} observations. However, the ratio of the period excess from \textit{TESS} data and the period deficit observed by \citet{2013MNRAS.435..707A} is surprisingly close to the 7/4 value (Fig.~\ref{fig:eps_pm}) predicted by \citet{2013PASJ...65...95O}. This could support the idea that the negative superhump observed by \citet{2013MNRAS.435..707A} was an ordinary negative superhump, whilst during \textit{TESS} observations an impulsive negative superhump was present. Moreover, impulsive negative superhumps are only observed during a decline from a dwarf nova outburst, i.e. when the mass transfer rate through the accretion disc is decreasing. Since the brightness of AQ~Men was decreasing during the period covered by \textit{TESS} (Fig.~\ref{fig:ASASSN}), a similar decrease in mass transfer rate trough the accretion disc occurred in AQ~Men. However, the mass transfer rate decrease in AQ~Men was orders of magnitude smaller compared to a dwarf nova and it is not clear whether it could result in an impulsive negative superhump.

An alternative explanation of why the period excess and period deficit ratio from \textit{TESS} data does not fit the theoretical prediction is that the signal observed at the frequency $\omega_-$ was wrongly interpreted as a negative superhump. Instead, an alternative explanation could be that the depth of eclipses in AQ~Men is changing with the superorbital frequency (Fig.~\ref{fig:Orb_change_superorb}). In signal processing, this can be interpreted as a modulation of an amplitude of the orbital signal with the modulation frequency equal to the disc precession frequency. Such a modulation is known to manifest itself in power spectra with side peaks, that in the case of AQ~Men would be located at frequencies $\omega_0-N$ and $\omega_0+N$ \citep[e.g.][]{2011MNRAS.417..974B}. Therefore, the signal at frequency $\omega_-=\omega_0+\mathrm{N}$ can be interpreted as a side peak caused by modulation of orbital variability amplitude rather than a negative superhump caused by a shifting hotspot around the face of a tilted, retrogradely precessing disc. The modulation at $\omega_0-\mathrm{N}$ in this case would be blended with a genuine positive superhump. However, we dismiss this interpretation as the change of eclipse depths is a relatively small effect (Fig.~\ref{fig:Orb_change_superorb}) and is not likely to produce such a strong artificial signal (Table~\ref{tab:table_frq}). In order to test that we created an artificial \textit{TESS} light curve of AQ~Men with only the orbital variability present. Periodogram of this artificial light curve showed that the changes in eclipse depths observed in AQ~Men could not result in an artificial signals strong enough to be detected with our accuracy.

The interpretation of the change in negative superhump period excess that we favour is the change of accretion disc inclination. It was shown that the negative superhump period excess is a function of both the disc inclination angle as well as the binary mass ratio (eq. 54-56 in \citealt{2009ApJ...705..603M}), which is not taken into account in relations such as the one presented in Fig.~\ref{fig:Porb_eps}. In particular, it is possible that the accretion disc was depleted during the low state observed in ASAS-SN data and it was reshaped at a different inclination angle. On other hand, the positive superhump is not affected by the accretion disc inclination angle and is following the standard relation (Fig.~\ref{fig:Porb_eps}).  Moreover, eq. 54 of \citet{2009ApJ...705..603M} suggests that the accretion disc in AQ~Men was more out of the orbital plane in observations of \citet{2013MNRAS.435..707A} than during the \textit{TESS} observations. It may be hypothesised that during \citet{2013MNRAS.435..707A} observations the accretion disc was so far out of the orbital plane that it was not affected by the 3:1 resonance, and because of that the positive superhump was not excited at that time. The change in the accretion disc inclination could also explain the shift in phase of the orbital hotspot between the \textit{TESS}  and \citet{2013MNRAS.435..707A} observations. The change of accretion disc inclination implies a shift of the position at which the accretion stream is hitting the accretion disc, and therefore implies a shift in the observed hotspot hump phase. Alternatively, the lack of positive superhump in observations of \citet{2013MNRAS.435..707A} may be explained if the accretion disc was smaller during their observations and it did not extend to the 3:1 resonance radius. This would be supported by the fact that AQ~Men shows high variability in mass transfer rate (Fig.~\ref{fig:ASASSN}).

\section{Summary}\label{sec:conclusions}

In this work we analysed \textit{TESS} observations of a nova-like variable AQ~Men which provided the most sensitive and longest monitoring of AQ~Men to date. The main findings of the work include:

\begin{itemize}
\item AQ~Men is a nova-like variable that experienced a low state at  MJD$\simeq$56970 that lasted $\sim$100~days. 
\item The negative superhump frequency and superorbital frequency changed significantly compared to the observations of \citet{2013MNRAS.435..707A}. The negative superhump shape changed slightly compared to past observations.
\item The positive superhump was detected for the first time in AQ~Men. The positive superhump shape is strongly non-sinusoidal, which is not expected for a nova-like variable.
\item The superorbital frequency increased between the \citet{2013MNRAS.435..707A} and \textit{TESS} observations. However, across the \textit{TESS} observations the superorbital frequency was decreasing, implying that the rate of frequency change inverted at least once in the past. The rate of change of the positive superhump frequency was consistent with a negative value of the superorbital frequency rate of change. This behaviour was accompanied by a decrease in mass transfer rate in the system.
\item The amplitudes of the orbital and superorbital variabilities were variable and seemed correlated most of the time. The exception was at the start of \textit{TESS} observations, where the amplitude of the superorbital variability was the highest during the period covered by \textit{TESS}, while the amplitude of the orbital variability at this time was close to the mean value.  Moreover, at the  start of \textit{TESS} observations the shape of superorbital variability was significantly non-sinusoidal, contrary to later observations.
\item In the measurements of the depths of eclipses, the superorbital frequency was detected. The eclipses were deepest at the superorbital phase $\sim$0.75, while at the superorbital phase $\sim$0.0 eclipses were shallowest. Contrary to the eclipse depth, the amplitude of the hotspot hump was not affected by the superorbital variability.  
\end{itemize}

The results show that AQ~Men is a useful tool in studying accretion discs in cataclysmic variables. In particular, the changes of eclipse depths with the superorbital frequency were predicted based on the assumption that the superorbital frequency is due to a precessing accretion disc that is out of the orbital plane. As the nature of negative superhumps and the superorbital signals are still under discussion, the detection of such changes serves as a test of the nature of the two signals. Because of this, AQ~Men should be an object of light curve modelling in the future, and in this way it can serve as a tool to probe the accretion disc structure and further test tilted accretion disc models. The \textit{TESS} observations covered only a period of a small, consistent decrease in mass transfer rate in AQ~Men. Since both the superhump and superorbital signals showed apparent correlation with changes in the mass transfer rate, the system should be monitored in the future to study behaviour of superhump and superorbital signals when further mass transfer rate variability is present.

\section*{Acknowledgements}

This  study  has  been  supported by the NASA Astrophysics Data Analysis Program 80NSSC20K0439. ND is supported by a Vidi grant from the Netherlands Organization for Scientific Research (NWO).

This paper includes data  collected  with  the TESS  mission,  obtained  from  the MAST  data  archive  at  the  Space  Telescope  Science  Institute (STScI). Funding for the TESS mission is provided by the NASA Explorer Program. STScI is operated by the Association  of  Universities  for  Research  in  Astronomy,  Inc.,under NASA contract NAS 5-26555.

\section*{DATA AVAILABILITY}
The data underlying this article were accessed from the Mikulski Archive for Space Telescope (MAST\footnote{http://archive.stsci.edu/}). The derived data generated in this research will be shared on reasonable request to the corresponding author.





\bibliographystyle{mnras}
\bibliography{literature} 

\begin{thebibliography}{}
\makeatletter
\relax
\def\mn@urlcharsother{\let\do\@makeother \do\$\do\&\do\#\do\^\do\_\do\%\do\~}
\def\mn@doi{\begingroup\mn@urlcharsother \@ifnextchar [ {\mn@doi@}
  {\mn@doi@[]}}
\def\mn@doi@[#1]#2{\def\@tempa{#1}\ifx\@tempa\@empty \href
  {http://dx.doi.org/#2} {doi:#2}\else \href {http://dx.doi.org/#2} {#1}\fi
  \endgroup}
\def\mn@eprint#1#2{\mn@eprint@#1:#2::\@nil}
\def\mn@eprint@arXiv#1{\href {http://arxiv.org/abs/#1} {{\tt arXiv:#1}}}
\def\mn@eprint@dblp#1{\href {http://dblp.uni-trier.de/rec/bibtex/#1.xml}
  {dblp:#1}}
\def\mn@eprint@#1:#2:#3:#4\@nil{\def\@tempa {#1}\def\@tempb {#2}\def\@tempc
  {#3}\ifx \@tempc \@empty \let \@tempc \@tempb \let \@tempb \@tempa \fi \ifx
  \@tempb \@empty \def\@tempb {arXiv}\fi \@ifundefined
  {mn@eprint@\@tempb}{\@tempb:\@tempc}{\expandafter \expandafter \csname
  mn@eprint@\@tempb\endcsname \expandafter{\@tempc}}}

\bibitem[\protect\citeauthoryear{{Andronov} et~al.,}{{Andronov}
  et~al.}{2003}]{2003ASPC..292..313A}
{Andronov} I.~L.,  et~al., 2003, in {Sterken} C.,  ed.,  Astronomical Society
  of the Pacific Conference Series Vol. 292, Interplay of Periodic, Cyclic and
  Stochastic Variability in Selected Areas of the H-R Diagram. p.~313

\bibitem[\protect\citeauthoryear{{Armstrong} et~al.,}{{Armstrong}
  et~al.}{2013}]{2013MNRAS.435..707A}
{Armstrong} E.,  et~al., 2013, \mn@doi [\mnras] {10.1093/mnras/stt1335}, \href
  {https://ui.adsabs.harvard.edu/abs/2013MNRAS.435..707A} {435, 707}

\bibitem[\protect\citeauthoryear{{Astropy Collaboration} et~al.,}{{Astropy
  Collaboration} et~al.}{2013}]{astropy:2013}
{Astropy Collaboration} et~al., 2013, \mn@doi [\aap]
  {10.1051/0004-6361/201322068}, \href
  {http://adsabs.harvard.edu/abs/2013A%26A...558A..33A} {558, A33}

\bibitem[\protect\citeauthoryear{{Astropy Collaboration} et~al.,}{{Astropy
  Collaboration} et~al.}{2018}]{astropy:2018}
{Astropy Collaboration} et~al., 2018, \mn@doi [aj] {10.3847/1538-3881/aabc4f},
  \href {https://ui.adsabs.harvard.edu/abs/2018AJ....156..123A} {156, 123}

\bibitem[\protect\citeauthoryear{{Belova}, {Suleimanov}, {Bikmaev}, {Khamitov},
  {Zhukov}, {Senio}, {Belov}  \& {Sakhibullin}}{{Belova}
  et~al.}{2013}]{2013AstL...39..111B}
{Belova} A.~I.,  {Suleimanov} V.~F.,  {Bikmaev} I.~F.,  {Khamitov} I.~M.,
  {Zhukov} G.~V.,  {Senio} D.~S.,  {Belov} I.~Y.,   {Sakhibullin} N.~A.,  2013,
  \mn@doi [Astronomy Letters] {10.1134/S1063773713020011}, \href
  {https://ui.adsabs.harvard.edu/abs/2013AstL...39..111B} {39, 111}

\bibitem[\protect\citeauthoryear{{Benk{\H{o}}}, {Szab{\'o}}  \&
  {Papar{\'o}}}{{Benk{\H{o}}} et~al.}{2011}]{2011MNRAS.417..974B}
{Benk{\H{o}}} J.~M.,  {Szab{\'o}} R.,   {Papar{\'o}} M.,  2011, \mn@doi
  [\mnras] {10.1111/j.1365-2966.2011.19313.x}, \href
  {https://ui.adsabs.harvard.edu/abs/2011MNRAS.417..974B} {417, 974}

\bibitem[\protect\citeauthoryear{{Bruch}}{{Bruch}}{2020}]{2020NewA...7801369B}
{Bruch} A.,  2020, \mn@doi [\na] {10.1016/j.newast.2020.101369}, \href
  {https://ui.adsabs.harvard.edu/abs/2020NewA...7801369B} {78, 101369}

\bibitem[\protect\citeauthoryear{{Chen}, {O'Donoghue}, {Stobie}, {Kilkenny}  \&
  {Warner}}{{Chen} et~al.}{2001}]{2001MNRAS.325...89C}
{Chen} A.,  {O'Donoghue} D.,  {Stobie} R.~S.,  {Kilkenny} D.,   {Warner} B.,
  2001, \mn@doi [\mnras] {10.1046/j.1365-8711.2001.04322.x}, \href
  {https://ui.adsabs.harvard.edu/abs/2001MNRAS.325...89C} {325, 89}

\bibitem[\protect\citeauthoryear{{Court} et~al.,}{{Court}
  et~al.}{2019}]{2019MNRAS.488.4149C}
{Court} J.~M.~C.,  et~al., 2019, \mn@doi [\mnras] {10.1093/mnras/stz2015},
  \href {https://ui.adsabs.harvard.edu/abs/2019MNRAS.488.4149C} {488, 4149}

\bibitem[\protect\citeauthoryear{{Dorn-Wallenstein}, {Levesque}  \&
  {Davenport}}{{Dorn-Wallenstein} et~al.}{2019}]{2019ApJ...878..155D}
{Dorn-Wallenstein} T.~Z.,  {Levesque} E.~M.,   {Davenport} J. R.~A.,  2019,
  \mn@doi [\apj] {10.3847/1538-4357/ab223f}, \href
  {https://ui.adsabs.harvard.edu/abs/2019ApJ...878..155D} {878, 155}

\bibitem[\protect\citeauthoryear{{Fontaine} et~al.,}{{Fontaine}
  et~al.}{2011}]{2011ApJ...726...92Fs}
{Fontaine} G.,  et~al., 2011, \mn@doi [\apj] {10.1088/0004-637X/726/2/92},
  \href {https://ui.adsabs.harvard.edu/abs/2011ApJ...726...92F} {726, 92}

\bibitem[\protect\citeauthoryear{{Godon}, {Sion}, {Barrett}  \&
  {Szkody}}{{Godon} et~al.}{2009}]{2009ApJ...701.1091G}
{Godon} P.,  {Sion} E.~M.,  {Barrett} P.~E.,   {Szkody} P.,  2009, \mn@doi
  [\apj] {10.1088/0004-637X/701/2/1091}, \href
  {https://ui.adsabs.harvard.edu/abs/2009ApJ...701.1091G} {701, 1091}

\bibitem[\protect\citeauthoryear{{Hirose} \& {Osaki}}{{Hirose} \&
  {Osaki}}{1990}]{1990PASJ...42..135H}
{Hirose} M.,  {Osaki} Y.,  1990, \pasj, \href
  {https://ui.adsabs.harvard.edu/abs/1990PASJ...42..135H} {42, 135}

\bibitem[\protect\citeauthoryear{{Jenkins} et~al.,}{{Jenkins}
  et~al.}{2016}]{2016SPIE.9913E..3EJ}
{Jenkins} J.~M.,  et~al., 2016, in \procspie. p. 99133E,
  \mn@doi{10.1117/12.2233418}

\bibitem[\protect\citeauthoryear{{Kato} \& {Hiroyuki}}{{Kato} \&
  {Hiroyuki}}{2013}]{2013PASJ...65...76K}
{Kato} T.,  {Hiroyuki} M.,  2013, \mn@doi [\pasj] {10.1093/pasj/65.4.76}, \href
  {https://ui.adsabs.harvard.edu/abs/2013PASJ...65...76K} {65, 76}

\bibitem[\protect\citeauthoryear{{Kato} et~al.,}{{Kato}
  et~al.}{2017}]{2017PASJ...69...75K}
{Kato} T.,  et~al., 2017, \mn@doi [\pasj] {10.1093/pasj/psx058}, \href
  {https://ui.adsabs.harvard.edu/abs/2017PASJ...69...75K} {69, 75}

\bibitem[\protect\citeauthoryear{{Kato} et~al.,}{{Kato}
  et~al.}{2020}]{2020PASJ...72...14K}
{Kato} T.,  et~al., 2020, \mn@doi [\pasj] {10.1093/pasj/psz134}, \href
  {https://ui.adsabs.harvard.edu/abs/2020PASJ...72...14K} {72, 14}

\bibitem[\protect\citeauthoryear{{Khruzina}, {Katysheva}, {Golysheva}  \&
  {Shugarov}}{{Khruzina} et~al.}{2015}]{2015EAS....71..149K}
{Khruzina} T.,  {Katysheva} N.,  {Golysheva} P.,   {Shugarov} S.,  2015, in EAS
  Publications Series. pp 149--150, \mn@doi{10.1051/eas/1571032}

\bibitem[\protect\citeauthoryear{{Kim}, {Andronov}, {Cha}, {Chinarova}  \&
  {Yoon}}{{Kim} et~al.}{2009}]{2009A&A...496..765K}
{Kim} Y.,  {Andronov} I.~L.,  {Cha} S.~M.,  {Chinarova} L.~L.,   {Yoon} J.~N.,
  2009, \mn@doi [\aap] {10.1051/0004-6361:200810005}, \href
  {https://ui.adsabs.harvard.edu/abs/2009A&A...496..765K} {496, 765}

\bibitem[\protect\citeauthoryear{{Knigge}, {Baraffe}  \& {Patterson}}{{Knigge}
  et~al.}{2011}]{2011ApJS..194...28K}
{Knigge} C.,  {Baraffe} I.,   {Patterson} J.,  2011, \mn@doi [\apjs]
  {10.1088/0067-0049/194/2/28}, \href
  {https://ui.adsabs.harvard.edu/abs/2011ApJS..194...28K} {194, 28}

\bibitem[\protect\citeauthoryear{{Kochanek} et~al.,}{{Kochanek}
  et~al.}{2017}]{2017PASP..129j4502K}
{Kochanek} C.~S.,  et~al., 2017, \mn@doi [\pasp] {10.1088/1538-3873/aa80d9},
  \href {https://ui.adsabs.harvard.edu/abs/2017PASP..129j4502K} {129, 104502}

\bibitem[\protect\citeauthoryear{{Kraicheva}, {Stanishev}, {Genkov}  \&
  {Iliev}}{{Kraicheva} et~al.}{1999}]{1999A&A...351..607K}
{Kraicheva} Z.,  {Stanishev} V.,  {Genkov} V.,   {Iliev} L.,  1999, \aap, \href
  {https://ui.adsabs.harvard.edu/abs/1999A&A...351..607K} {351, 607}

\bibitem[\protect\citeauthoryear{{Larwood}}{{Larwood}}{1998}]{1998MNRAS.299L..32L}
{Larwood} J.,  1998, \mn@doi [\mnras] {10.1046/j.1365-8711.1998.01978.x}, \href
  {https://ui.adsabs.harvard.edu/abs/1998MNRAS.299L..32L} {299, L32}

\bibitem[\protect\citeauthoryear{{Lomb}}{{Lomb}}{1976}]{1976Ap&SS..39..447L}
{Lomb} N.~R.,  1976, \mn@doi [\apss] {10.1007/BF00648343}, \href
  {https://ui.adsabs.harvard.edu/abs/1976Ap&SS..39..447L} {39, 447}

\bibitem[\protect\citeauthoryear{{Montgomery}}{{Montgomery}}{2001}]{2001MNRAS.325..761M}
{Montgomery} M.~M.,  2001, \mn@doi [\mnras] {10.1046/j.1365-8711.2001.04487.x},
  \href {https://ui.adsabs.harvard.edu/abs/2001MNRAS.325..761M} {325, 761}

\bibitem[\protect\citeauthoryear{{Montgomery}}{{Montgomery}}{2009}]{2009ApJ...705..603M}
{Montgomery} M.~M.,  2009, \mn@doi [\apj] {10.1088/0004-637X/705/1/603}, \href
  {https://ui.adsabs.harvard.edu/abs/2009ApJ...705..603M} {705, 603}

\bibitem[\protect\citeauthoryear{{Montgomery}}{{Montgomery}}{2012}]{2012ApJ...745L..25M}
{Montgomery} M.~M.,  2012, \mn@doi [\apjl] {10.1088/2041-8205/745/2/L25}, \href
  {https://ui.adsabs.harvard.edu/abs/2012ApJ...745L..25M} {745, L25}

\bibitem[\protect\citeauthoryear{{Ohshima} et~al.,}{{Ohshima}
  et~al.}{2014}]{2014PASJ...66...67O}
{Ohshima} T.,  et~al., 2014, \mn@doi [\pasj] {10.1093/pasj/psu038}, \href
  {https://ui.adsabs.harvard.edu/abs/2014PASJ...66...67O} {66, 67}

\bibitem[\protect\citeauthoryear{{Osaki} \& {Kato}}{{Osaki} \&
  {Kato}}{2013}]{2013PASJ...65...95O}
{Osaki} Y.,  {Kato} T.,  2013, \mn@doi [\pasj] {10.1093/pasj/65.5.95}, \href
  {https://ui.adsabs.harvard.edu/abs/2013PASJ...65...95O} {65, 95}

\bibitem[\protect\citeauthoryear{{Papadaki}, {Boffin}, {Sterken}, {Stanishev},
  {Cuypers}, {Boumis}, {Akras}  \& {Alikakos}}{{Papadaki}
  et~al.}{2006}]{2006A&A...456..599P}
{Papadaki} C.,  {Boffin} H.~M.~J.,  {Sterken} C.,  {Stanishev} V.,  {Cuypers}
  J.,  {Boumis} P.,  {Akras} S.,   {Alikakos} J.,  2006, \mn@doi [\aap]
  {10.1051/0004-6361:20054679}, \href
  {https://ui.adsabs.harvard.edu/abs/2006A&A...456..599P} {456, 599}

\bibitem[\protect\citeauthoryear{{Patterson}}{{Patterson}}{1995}]{1995PASP..107..657P}
{Patterson} J.,  1995, \mn@doi [\pasp] {10.1086/133605}, \href
  {https://ui.adsabs.harvard.edu/abs/1995PASP..107..657P} {107, 657}

\bibitem[\protect\citeauthoryear{{Patterson}}{{Patterson}}{1999}]{1999dicb.conf...61P}
{Patterson} J.,  1999, in {Mineshige} S.,  {Wheeler} J.~C.,  eds, Disk
  Instabilities in Close Binary Systems. p.~61

\bibitem[\protect\citeauthoryear{{Patterson}, {Kemp}, {Saad}, {Skillman},
  {Harvey}, {Fried}, {Thorstensen}  \& {Ashley}}{{Patterson}
  et~al.}{1997}]{1997PASP..109..468P}
{Patterson} J.,  {Kemp} J.,  {Saad} J.,  {Skillman} D.~R.,  {Harvey} D.,
  {Fried} R.,  {Thorstensen} J.~R.,   {Ashley} R.,  1997, \mn@doi [\pasp]
  {10.1086/133903}, \href
  {https://ui.adsabs.harvard.edu/abs/1997PASP..109..468P} {109, 468}

\bibitem[\protect\citeauthoryear{{Patterson} et~al.,}{{Patterson}
  et~al.}{2005}]{2005PASP..117.1204P}
{Patterson} J.,  et~al., 2005, \mn@doi [\pasp] {10.1086/447771}, \href
  {https://ui.adsabs.harvard.edu/abs/2005PASP..117.1204P} {117, 1204}

\bibitem[\protect\citeauthoryear{Retter, Chou, Bedding  \& Naylor}{Retter
  et~al.}{2002}]{10.1046/j.1365-8711.2002.05280.x}
Retter A.,  Chou Y.,  Bedding T.~R.,   Naylor T.,  2002, \mn@doi [Monthly
  Notices of the Royal Astronomical Society]
  {10.1046/j.1365-8711.2002.05280.x}, 330, L37

\bibitem[\protect\citeauthoryear{{Ricker} et~al.,}{{Ricker}
  et~al.}{2015}]{2015JATIS...1a4003R}
{Ricker} G.~R.,  et~al., 2015, \mn@doi [Journal of Astronomical Telescopes,
  Instruments, and Systems] {10.1117/1.JATIS.1.1.014003}, \href
  {https://ui.adsabs.harvard.edu/abs/2015JATIS...1a4003R} {1, 014003}

\bibitem[\protect\citeauthoryear{{Ringwald}, {Velasco}, {Roveto}  \&
  {Meyers}}{{Ringwald} et~al.}{2012}]{2012NewA...17..433R}
{Ringwald} F.~A.,  {Velasco} K.,  {Roveto} J.~J.,   {Meyers} M.~E.,  2012,
  \mn@doi [\na] {10.1016/j.newast.2011.11.007}, \href
  {https://ui.adsabs.harvard.edu/abs/2012NewA...17..433R} {17, 433}

\bibitem[\protect\citeauthoryear{{Rolfe}, {Haswell}  \& {Patterson}}{{Rolfe}
  et~al.}{2001}]{2001MNRAS.324..529R}
{Rolfe} D.~J.,  {Haswell} C.~A.,   {Patterson} J.,  2001, \mn@doi [\mnras]
  {10.1046/j.1365-8711.2001.04248.x}, \href
  {https://ui.adsabs.harvard.edu/abs/2001MNRAS.324..529R} {324, 529}

\bibitem[\protect\citeauthoryear{{Rutkowski}, {Ak}, {Marsh}  \&
  {Eker}}{{Rutkowski} et~al.}{2013}]{2013AcA....63..225R}
{Rutkowski} A.,  {Ak} T.,  {Marsh} T.~R.,   {Eker} Z.,  2013, \actaa, \href
  {https://ui.adsabs.harvard.edu/abs/2013AcA....63..225R} {63, 225}

\bibitem[\protect\citeauthoryear{{Scargle}}{{Scargle}}{1982}]{1982ApJ...263..835S}
{Scargle} J.~D.,  1982, \mn@doi [\apj] {10.1086/160554}, \href
  {https://ui.adsabs.harvard.edu/abs/1982ApJ...263..835S} {263, 835}

\bibitem[\protect\citeauthoryear{{Schmidtobreick}, {Rodr{\'\i}guez-Gil}  \&
  {G{\"a}nsicke}}{{Schmidtobreick} et~al.}{2012}]{2012MmSAI..83..610S}
{Schmidtobreick} L.,  {Rodr{\'\i}guez-Gil} P.,   {G{\"a}nsicke} B.~T.,  2012,
  \memsai, \href {https://ui.adsabs.harvard.edu/abs/2012MmSAI..83..610S} {83,
  610}

\bibitem[\protect\citeauthoryear{{Shappee} et~al.,}{{Shappee}
  et~al.}{2014}]{2014ApJ...788...48S}
{Shappee} B.~J.,  et~al., 2014, \mn@doi [\apj] {10.1088/0004-637X/788/1/48},
  \href {https://ui.adsabs.harvard.edu/abs/2014ApJ...788...48S} {788, 48}

\bibitem[\protect\citeauthoryear{{Simpson} \& {Wood}}{{Simpson} \&
  {Wood}}{1998}]{1998ApJ...506..360S}
{Simpson} J.~C.,  {Wood} M.~A.,  1998, \mn@doi [\apj] {10.1086/306221}, \href
  {https://ui.adsabs.harvard.edu/abs/1998ApJ...506..360S} {506, 360}

\bibitem[\protect\citeauthoryear{{Skillman} et~al.,}{{Skillman}
  et~al.}{1998}]{1998ApJ...503L..67S}
{Skillman} D.~R.,  et~al., 1998, \mn@doi [\apjl] {10.1086/311534}, \href
  {https://ui.adsabs.harvard.edu/abs/1998ApJ...503L..67S} {503, L67}

\bibitem[\protect\citeauthoryear{{Skillman}, {Patterson}, {Kemp}, {Harvey},
  {Fried}, {Retter}, {Lipkin}  \& {Vanmunster}}{{Skillman}
  et~al.}{1999}]{1999PASP..111.1281S}
{Skillman} D.~R.,  {Patterson} J.,  {Kemp} J.,  {Harvey} D.~A.,  {Fried} R.~E.,
   {Retter} A.,  {Lipkin} Y.,   {Vanmunster} T.,  1999, \mn@doi [\pasp]
  {10.1086/316437}, \href
  {https://ui.adsabs.harvard.edu/abs/1999PASP..111.1281S} {111, 1281}

\bibitem[\protect\citeauthoryear{{Smak}}{{Smak}}{1994}]{1994AcA....44..257S}
{Smak} J.,  1994, \actaa, \href
  {https://ui.adsabs.harvard.edu/abs/1994AcA....44..257S} {44, 257}

\bibitem[\protect\citeauthoryear{{Smak} \& {Stepien}}{{Smak} \&
  {Stepien}}{1975}]{1975AcA....25..379S}
{Smak} J.,  {Stepien} K.,  1975, \actaa, \href
  {https://ui.adsabs.harvard.edu/abs/1975AcA....25..379S} {25, 379}

\bibitem[\protect\citeauthoryear{{Stobie}, {Kilkenny}  \&
  {O'Donoghue}}{{Stobie} et~al.}{1995}]{1995Ap&SS.230..101S}
{Stobie} R.~S.,  {Kilkenny} D.,   {O'Donoghue} D.,  1995, \mn@doi [\apss]
  {10.1007/BF00658172}, \href
  {https://ui.adsabs.harvard.edu/abs/1995Ap&SS.230..101S} {230, 101}

\bibitem[\protect\citeauthoryear{{Tampo} et~al.,}{{Tampo}
  et~al.}{2020}]{2020PASJ...72...49T}
{Tampo} Y.,  et~al., 2020, \mn@doi [\pasj] {10.1093/pasj/psaa043}, \href
  {https://ui.adsabs.harvard.edu/abs/2020PASJ...72...49T} {72, 49}

\bibitem[\protect\citeauthoryear{{Thomas} \& {Wood}}{{Thomas} \&
  {Wood}}{2015}]{2015ApJ...803...55T}
{Thomas} D.~M.,  {Wood} M.~A.,  2015, \mn@doi [\apj]
  {10.1088/0004-637X/803/2/55}, \href
  {https://ui.adsabs.harvard.edu/abs/2015ApJ...803...55T} {803, 55}

\bibitem[\protect\citeauthoryear{{Warner}}{{Warner}}{2003}]{2003cvs..book.....W}
{Warner} B.,  2003, {Cataclysmic Variable Stars},
  \mn@doi{10.1017/CBO9780511586491.
}

\bibitem[\protect\citeauthoryear{{Weingrill}, {Kleinschuster}, {Kuschnig},
  {Matthews}, {Moffat}, {Rucinski}, {Sasselov}  \& {Weiss}}{{Weingrill}
  et~al.}{2009}]{2009CoAst.159..114W}
{Weingrill} J.,  {Kleinschuster} G.,  {Kuschnig} R.,  {Matthews} J.~M.,
  {Moffat} A.,  {Rucinski} S.,  {Sasselov} D.,   {Weiss} W.~W.,  2009, \mn@doi
  [Communications in Asteroseismology] {10.1553/cia159s114}, \href
  {https://ui.adsabs.harvard.edu/abs/2009CoAst.159..114W} {159, 114}

\bibitem[\protect\citeauthoryear{{Whitehurst}}{{Whitehurst}}{1988a}]{1988MNRAS.232...35W}
{Whitehurst} R.,  1988a, \mn@doi [\mnras] {10.1093/mnras/232.1.35}, \href
  {https://ui.adsabs.harvard.edu/abs/1988MNRAS.232...35W} {232, 35}

\bibitem[\protect\citeauthoryear{{Whitehurst}}{{Whitehurst}}{1988b}]{1988MNRAS.233..529W}
{Whitehurst} R.,  1988b, \mn@doi [\mnras] {10.1093/mnras/233.3.529}, \href
  {https://ui.adsabs.harvard.edu/abs/1988MNRAS.233..529W} {233, 529}

\bibitem[\protect\citeauthoryear{{Wood}, {Montgomery}  \& {Simpson}}{{Wood}
  et~al.}{2000}]{2000ApJ...535L..39W}
{Wood} M.~A.,  {Montgomery} M.~M.,   {Simpson} J.~C.,  2000, \mn@doi [\apjl]
  {10.1086/312687}, \href
  {https://ui.adsabs.harvard.edu/abs/2000ApJ...535L..39W} {535, L39}

\bibitem[\protect\citeauthoryear{{Wood}, {Thomas}  \& {Simpson}}{{Wood}
  et~al.}{2009}]{2009MNRAS.398.2110W}
{Wood} M.~A.,  {Thomas} D.~M.,   {Simpson} J.~C.,  2009, \mn@doi [\mnras]
  {10.1111/j.1365-2966.2009.15252.x}, \href
  {https://ui.adsabs.harvard.edu/abs/2009MNRAS.398.2110W} {398, 2110}

\bibitem[\protect\citeauthoryear{{Wood}, {Still}, {Howell}, {Cannizzo}  \&
  {Smale}}{{Wood} et~al.}{2011}]{2011ApJ...741..105W}
{Wood} M.~A.,  {Still} M.~D.,  {Howell} S.~B.,  {Cannizzo} J.~K.,   {Smale}
  A.~P.,  2011, \mn@doi [\apj] {10.1088/0004-637X/741/2/105}, \href
  {https://ui.adsabs.harvard.edu/abs/2011ApJ...741..105W} {741, 105}

\makeatother
\end{thebibliography}



\appendix

\bsp	
\label{lastpage}
\end{document}